\documentclass[apj,numberedappendix,revtex4]{emulateapj}
\usepackage{color}
\definecolor{Mygreen}{rgb}{0.00, 0.5, 0.5}
\definecolor{Mypink}{rgb}{1.0, 0.0, 0.5}
\definecolor{Myblue}{rgb}{0.00, 0.2, 0.8}
\definecolor{Myred}{rgb}{0.80, 0.2, 0.0}
\usepackage[backref,breaklinks,colorlinks,citecolor=Myblue,linkcolor=Mygreen,urlcolor=Mygreen]{hyperref}
\usepackage{amssymb,amsmath}
\usepackage[mathscr]{eucal}
\usepackage{graphicx}
\usepackage{appendix}
\usepackage{natbib}
\usepackage{enumerate}
\usepackage{multirow}
\usepackage{bm}

\def\simlt{\lower.5ex\hbox{$\; \buildrel < \over \sim \;$}}
\def\simgt{\lower.5ex\hbox{$\; \buildrel > \over \sim \;$}}

\renewcommand{\arraystretch}{1.3}

\altaffiltext{\MIT}{Kavli Institute for Astrophysics and Space Research, Massachusetts Institute of Technology, 77 Massachusetts Avenue, Cambridge, MA 02139}
\altaffiltext{\UMis}{Department of Physics and Astronomy, University of Missouri, 5110 Rockhill Road, Kansas City, MO 64110, USA}
\altaffiltext{\LLR}{Laboratoire Leprince-Ringuet, \'Ecole Polytechnique, CNRS/IN2P3, 91128 Palaiseau, France}
\altaffiltext{\Cardiff} {Astronomy Instrumentation Group, University of Cardiff, UK}
\altaffiltext{\CEA1}{AIM, CEA, CNRS, Universit\'e Paris-Saclay, Universit\'e Paris Diderot, Sorbonne Paris Cit\'e, F-91191 Gif-sur-Yvette, France}
\altaffiltext{\IPAG}{Univ. Grenoble Alpes, CNRS, IPAG, 38000 Grenoble, France}
\altaffiltext{\IAS}{Institut d'Astrophysique Spatiale (IAS), CNRS and Universit\'e Paris Sud, Orsay, France}
\altaffiltext{\Neel}{Institut N\'eel, CNRS and Universit\'e Grenoble Alpes, France}
\altaffiltext{\LPSC}{Univ. Grenoble Alpes, CNRS, Grenoble INP, LPSC-IN2P3, 53, avenue des Martyrs, 38000 Grenoble, France}
\altaffiltext{\Roma}{Dipartimento di Fisica, Sapienza Universit\`a di Roma, Piazzale Aldo Moro 5, I-00185 Roma, Italy}
\altaffiltext{\IRAMF}{Institut de RadioAstronomie Millim\'etrique (IRAM), Grenoble, France}
\altaffiltext{\JPL}{Jet Propulsion Laboratory, California Institute of Technology, Pasadena, CA 91109, USA}
\altaffiltext{\CAB}{Centro de Astrobiolog\'ia (CSIC-INTA), Torrej\'on de Ardoz, 28850 Madrid, Spain}
\altaffiltext{\UFlo}{Department of Astronomy, University of Florida, 211 Bryant Space Center, Gainesville, FL 32611, USA}
\altaffiltext{\IRAME}{Institut de RadioAstronomie Millim\'etrique (IRAM), Granada, Spain}
\altaffiltext{\LAM}{Aix Marseille Universit\'e, CNRS, LAM (Laboratoire d'Astrophysique de Marseille) UMR 7326, 13388, Marseille, France}
\altaffiltext{\LERMA}{LERMA, Observatoire de Paris, PSL Research University, CNRS, Sorbonne Universit\'es, UPMC Univ., 75014, Paris, France}
\altaffiltext{\Arizona}{School of Earth and Space Exploration and Department of Physics, Arizona State University, Tempe, AZ 85287}
\altaffiltext{\IRAP}{IRAP, Universit\'e de Toulouse, CNRS, CNES, UPS, (Toulouse), France}
\altaffiltext{\IAP}{Institut d'Astrophysique de Paris, CNRS (UMR7095), 98 bis boulevard Arago, F-75014, Paris, France}
\altaffiltext{\UCal}{Department of Physics, University of California, One Shields Avenue, Davis, CA 95616, USA}

\def\MIT{1}
\def\UMis{2}
\def\LLR{3}
\def\Cardiff{4}
\def\CEA1{5}
\def\IPAG{6}
\def\IAS{7}
\def\Neel{8}
\def\LPSC{9}
\def\Roma{10}
\def\IRAMF{11}
\def\JPL{12}
\def\CAB{13}
\def\UFlo{14}
\def\IRAME{15}
\def\LAM{16}
\def\LERMA{17}
\def\Arizona{18}
\def\IRAP{19}
\def\IAP{20}
\def\UCal{21}

\begin{document}

\def\aj{AJ}%
\def\araa{ARA\&A}%
\def\apj{ApJ}%
\def\apjl{ApJL}%
\def\apjs{ApJS}%
\def\aap{Astron. Astrophys.}%
 \def\aapr{A\&A~Rev.}%
\def\aaps{A\&AS}%
\def\mnras{MNRAS}
\def\ssr{SSRv}
\def\nat{Nature}
\def\jcap{JCAP}

\def\Mgv{M_{\rm g,500}}
\def\Mg{M_{\rm g}}
\def\YX {Y_{\rm X}}
\def\LXv {L_{\rm X,500}}
\def\TX {T_{\rm X}}
\def\fgv {f_{\rm g,500}}
\def\fg  {f_{\rm g}}
\def\kT {{\rm k}T}
\def\ne {n_{\rm e}}
\def\Mv {M_{\rm 500}}
\def \Rv {R_{500}}
\def\keV {\rm keV}
\def\Yv{Y_{500}}

\def\MT {$M$--$T_{\rm X}$}
\def\MYX {$M$--$Y_{\rm X}$}
\def\MMg {$M_{500}$--$M_{\rm g,500}$}
\def\MgT {$M_{\rm g,500}$--$T_{\rm X}$}
\def\MgY {$M_{\rm g,500}$--$Y_{\rm X}$}

\def\msol {{\rm M_{\odot}}}

\def\lesssim{\mathrel{\hbox{\rlap{\hbox{\lower4pt\hbox{$\sim$}}}\hbox{$<$}}}}
\def\gtrsim{\mathrel{\hbox{\rlap{\hbox{\lower4pt\hbox{$\sim$}}}\hbox{$>$}}}}

\def\moo{MOO\,J1142$+$1527}

\def\xmm{XMM-{\it Newton}}
\def\planck{{\it Planck}} 
\def\chandra{{\it Chandra}}
\def \rosat {\hbox{\it ROSAT}}
\newcommand{\excpres}{{\gwpfont EXCPRES}}
\newcommand{\ma}[1]{\textcolor{red}{{ #1}}}
\title{Unveiling the merger dynamics of the most massive MaDCoWS cluster at $z = 1.2$ from a multi-wavelength mapping of its intracluster medium properties}


\author{
F.~Ruppin\altaffilmark{\MIT},
M. McDonald\altaffilmark{\MIT},
M. Brodwin\altaffilmark{\UMis},
R. Adam\altaffilmark{\LLR},
P. Ade\altaffilmark{\Cardiff},
P. Andr\'e\altaffilmark{\CEA1},
A. Andrianasolo\altaffilmark{\IPAG},
M. Arnaud\altaffilmark{\CEA1},
H. Aussel\altaffilmark{\CEA1},
I. Bartalucci\altaffilmark{\CEA1},
M. W. Bautz\altaffilmark{\MIT},
A. Beelen\altaffilmark{\IAS},
A. Beno\^it\altaffilmark{\Neel},
A. Bideaud\altaffilmark{\Neel},
O. Bourrion\altaffilmark{\LPSC},
M. Calvo\altaffilmark{\Neel},
A. Catalano\altaffilmark{\LPSC},
B. Comis\altaffilmark{\LPSC},
B. Decker\altaffilmark{\UMis},
M. De~Petris\altaffilmark{\Roma},
F.-X. D\'esert\altaffilmark{\IPAG},
S. Doyle\altaffilmark{\Cardiff},
E.~F.~C. Driessen\altaffilmark{\IRAMF},
P. R. M. Eisenhardt\altaffilmark{\JPL},
A. Gomez\altaffilmark{\CAB},
A. H. Gonzalez\altaffilmark{\UFlo},
J. Goupy\altaffilmark{\Neel},
F. K\'eruzor\'e\altaffilmark{\LPSC},
C. Kramer\altaffilmark{\IRAME},
B. Ladjelate\altaffilmark{\IRAME},
G. Lagache\altaffilmark{\LAM},
S. Leclercq\altaffilmark{\IRAMF},
J.-F. Lestrade\altaffilmark{\LERMA},
J.F. Mac\'ias-P\'erez\altaffilmark{\LPSC},
P. Mauskopf\altaffilmark{\Cardiff,\Arizona},
F. Mayet\altaffilmark{\LPSC},
A. Monfardini\altaffilmark{\Neel},
E. Moravec\altaffilmark{\UFlo},
L. Perotto\altaffilmark{\LPSC},
G. Pisano\altaffilmark{\Cardiff},
E. Pointecouteau\altaffilmark{\IRAP},
N. Ponthieu\altaffilmark{\IPAG},
G. W. Pratt\altaffilmark{\CEA1},
V. Rev\'eret\altaffilmark{\CEA1},
A. Ritacco\altaffilmark{\IRAME},
C. Romero\altaffilmark{\IRAMF},
H. Roussel\altaffilmark{\IAP},
K. Schuster\altaffilmark{\IRAMF},
S. Shu\altaffilmark{\IRAMF},
A. Sievers\altaffilmark{\IRAME},
S. A. Stanford\altaffilmark{\UCal},
D. Stern\altaffilmark{\JPL},
C. Tucker\altaffilmark{\Cardiff},
R. Zylka\altaffilmark{\IRAMF}
}

\email{Email: ruppin@mit.edu}

\begin{abstract}
The characterization of the Intra-Cluster Medium (ICM) properties of high-redshift galaxy clusters is fundamental to our understanding of large-scale structure formation processes. We present the results of a multi-wavelength analysis of the very massive cluster \moo\ at a redshift $z = 1.2$ discovered as part of the Massive and Distant Clusters of WISE Survey (MaDCoWS). This analysis is based on high angular resolution \chandra\ X-ray and NIKA2 Sunyaev-Zel'dovich (SZ) data. Although the X-ray data have only about 1700 counts, we are able to determine the ICM thermodynamic radial profiles, namely  temperature, entropy, and hydrostatic mass. These have been obtained with unprecedented precision at this redshift and up to $0.7R_{500}$, thanks to the combination of high-resolution X-ray and SZ data. The comparison between the galaxy distribution mapped in infrared by \emph{Spitzer} and the morphological properties of the ICM derived from the combined analysis of the \chandra\ and NIKA2 data leads us to the conclusion that the cluster is an on-going merger. We measure the hydrostatic mass profile of the cluster in four angular sectors centered on the large-scale X-ray centroid. This allows us to estimate a systematic uncertainty on the cluster total mass that characterizes both the impact of the observed deviations from spherical symmetry and of the core dynamics on the mass profile. We further combine the X-ray and SZ data at the pixel level to obtain maps of the temperature and entropy distributions averaged along the line of sight. We find a relatively low entropy core at the position of the X-ray peak and high temperature regions located on its south and west sides. The increase in ICM temperature at the location of the SZ peak is expected given the merger dynamics. This work demonstrates that the addition of spatially resolved SZ observations to low signal-to-noise X-ray data brings a high information gain on the characterization of the evolution of ICM thermodynamic properties at $z>1$.
\end{abstract}

 \vspace{-0.2in}

\section{Introduction}\label{sec:Introduction}
Clusters of galaxies form at the intersection of cosmic filaments and grow hierarchically through the joint process of slow accretion of surrounding material and merger events with sub-structures \citep[\emph{e.g.}][]{pre74}. Most of their baryonic matter content is made of a hot and diffuse plasma called the Intra-Cluster Medium (ICM) embedded within a dark matter halo. As the largest gravitationally bound objects, galaxy clusters provide a wealth of information on both the history of large scale structure formation and the dynamics of the Universe \citep[\emph{e.g.}][]{voi05}. The characterization of the evolution of the ICM thermodynamic properties with the mass and redshift of galaxy clusters thus gives us a way to test our current models describing the astrophysical processes that play a fundamental role during their growth as well as the underlying cosmology in which these processes take place.\\
\indent The vast majority of the detailed analyses of ICM astrophysical processes have been focused on $z<1$ clusters observed primarily in X-ray \citep[\emph{e.g.}][]{pra10,pac16,mcd19,cal19} and to some extent in SZ \citep[\emph{e.g.}][]{mro12,pla13}. These studies provided valuable constraints on a plethora of mechanisms such as gas cooling, feedback from active galactic nuclei, the physics behind cold and shock fronts, and merger dynamics. The comparison of both the amplitude and the shape of the mean radial distributions of ICM properties for representative cluster samples together with results from numerical simulations have enabled improving our understanding of cluster dynamics \citep[\emph{e.g.}][]{wal12,pik14}. Furthermore, knowing the mean ICM thermodynamic properties and how they scale with halo mass and redshift is essential to the use of clusters as cosmological probes \citep[\emph{e.g.}][]{pla16}.\\
\indent However, the most active epoch of cluster formation is assumed to lie at redshifts $1 < z < 2$ \citep[\emph{e.g.}][]{poo07}. During this period, the first extended protoclusters merged with galaxy groups and collapsed into more compact and massive halos \citep[\emph{e.g.}][]{mul15}. Also at these redshifts, cluster galaxies underwent a high star formation rate and an excess in the Active Galactic Nucleus (AGN) fraction is found \citep[\emph{e.g.}][]{alb16}. \begin{figure*}[t]
\centering
\includegraphics[height=6.4cm]{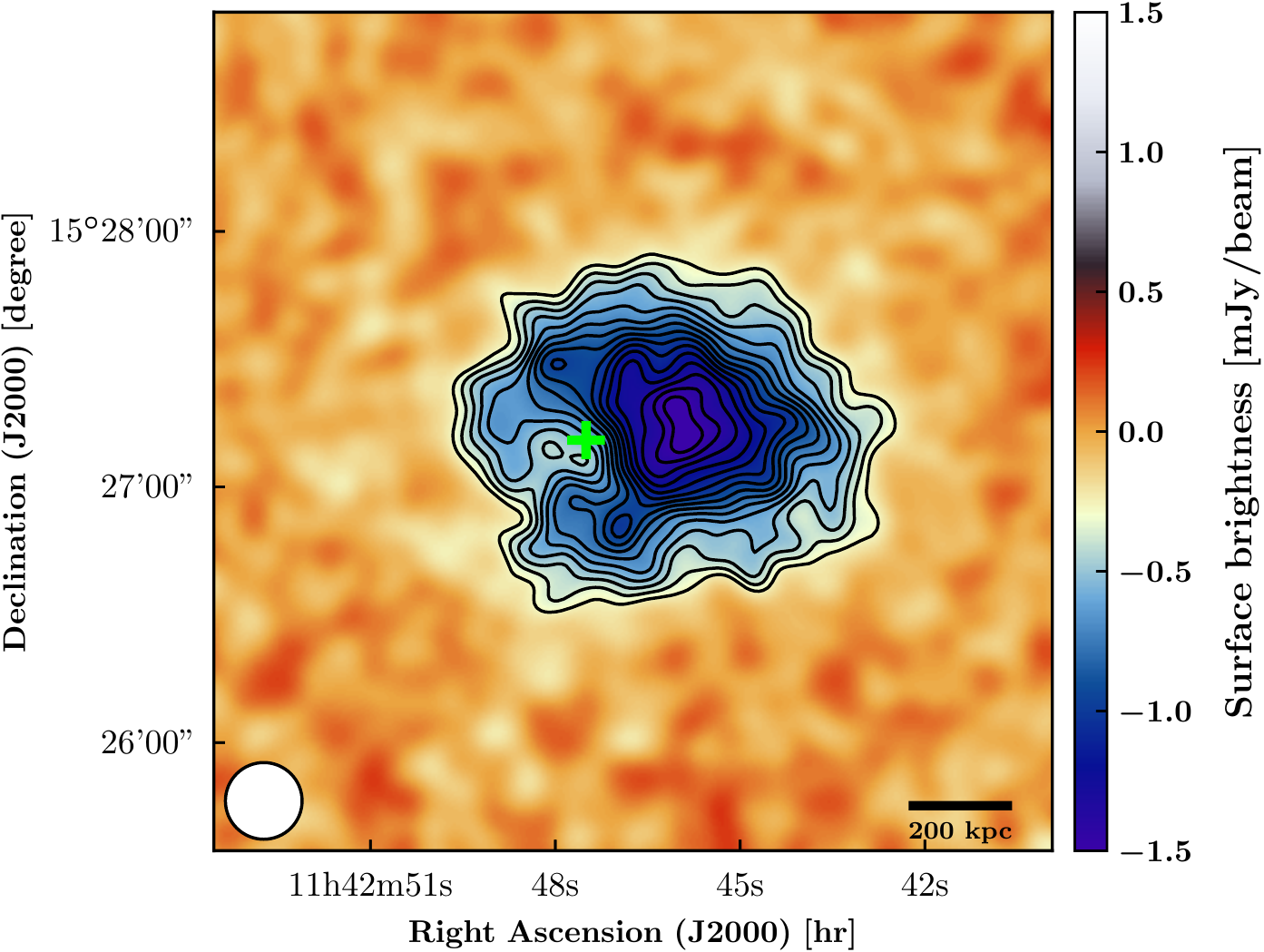}
\hspace{0.6cm}
\includegraphics[height=6.4cm]{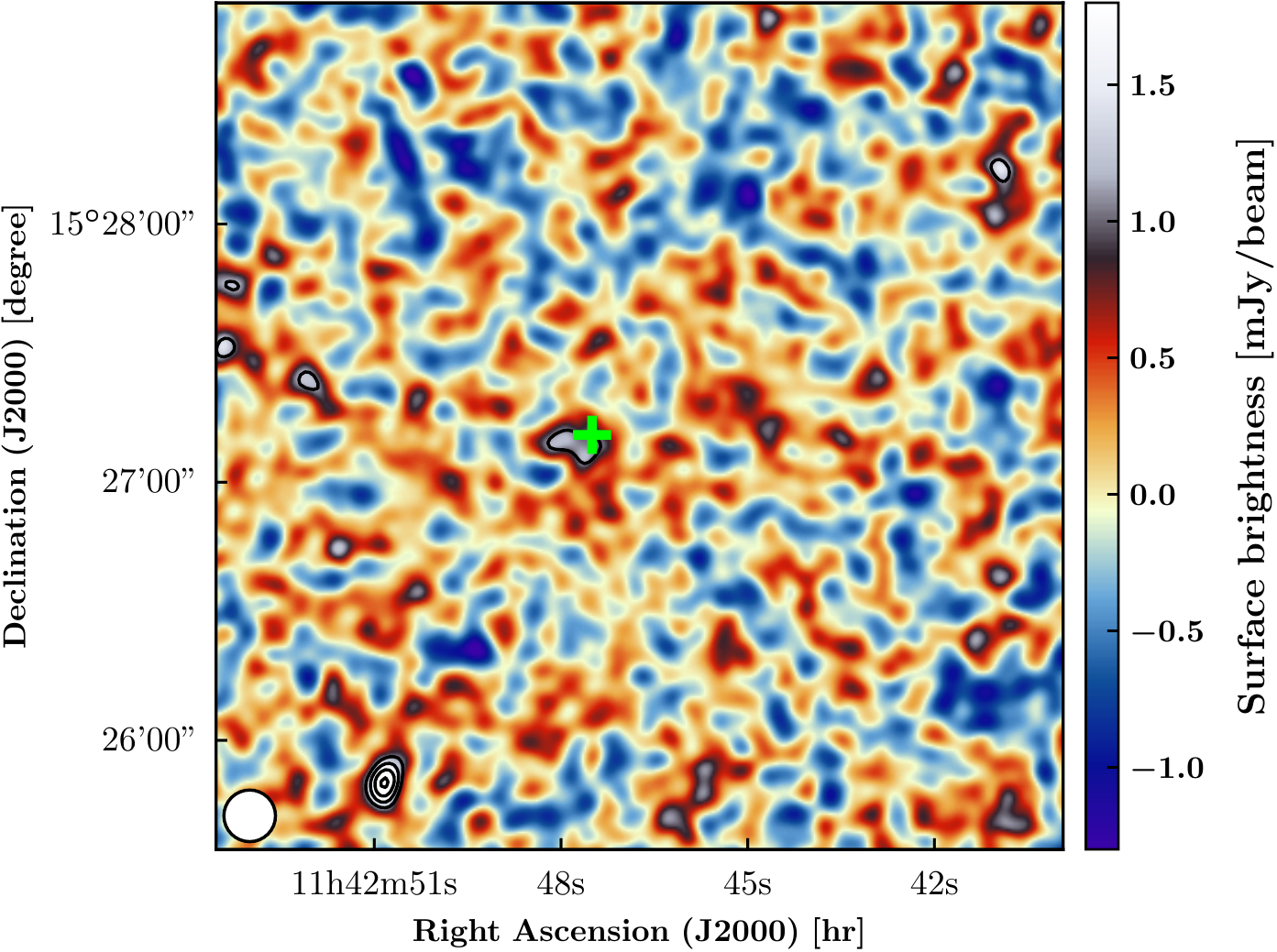}
\caption{{\footnotesize NIKA2 surface brightness maps of \moo\ at 150 GHz (left) and 260 GHz (right). The maps are smoothed with an additional 8 and 5~arcsec Full Width at Half Maximum (FWHM) Gaussian filter at 150 and 260 GHz respectively for display purposes. The significance contours in black start at $3\sigma$ with steps of $1\sigma$. We represent the width of the NIKA2 beams as white disks in the bottom left-hand corner of the maps. The considered FOV is 3.3 arcmin wide. The green cross in both panels gives the location of the radio source identified in the FIRST survey.}}
\label{fig:nk2_maps}
\end{figure*}Furthermore, the mean ICM thermodynamic properties found at low redshift may have significantly evolved since $z{\sim}2$ as the merger rate is expected to be much higher at $z\gtrsim 1$ \citep{mcd14}. This evolution may result in important modifications of the cosmological constraints issued from the study of cluster abundance \citep{rup19}. It is therefore essential to probe the ICM properties at redshifts higher than 1 to extend our knowledge of the distant progenitor objects of the most massive clusters at $z{\sim}0$.\\
\indent With the current X-ray observatories, the required exposures to probe the evolution of the radial distributions of all ICM thermodynamic properties at $z>1$ are extremely long and usually prevent such studies from being realized. Furthermore, X-ray selected samples may be affected by important selection biases \citep[\emph{e.g.}][]{eck11}. The past decade has seen the advent of large Sunyaev-Zel'dovich (SZ) surveys capable of detecting massive galaxy clusters up to high redshift. These include the \planck\ \citep{pla16b}, South Pole Telescope  \citep[SPT;][]{ble15}, and Atacama Cosmology Telescope \citep[ACT;][]{has13} surveys, with the latter two extending to $z>1$. However, the relatively low angular resolution of these instruments precludes these surveys from constraining the shape of the ICM pressure distribution or cluster morphology at high redshift. In contrast, the recently commissioned single-dish SZ instruments MUSTANG-2 \citep{dic14} and NIKA2 \citep{cal16,ada18,per19} open new possibilities regarding the characterization of the ICM properties even at the highest redshifts. Their high angular resolution (${\sim}9{-}18$ arcsec) and large field of view (${\sim}5{-}7$ arcmin) allows us to map the SZ signal over a range of scales that is similar to the X-ray measurements made by the current X-ray observatories such as \chandra\ and \xmm. Joint SZ and X-ray analyses represent a new frontier in high-redshift cluster studies, and promise to address outstanding questions associated with the formation and evolution of the ICM. At low redshift, such joint studies have yielded detailed characterization of individual cluster merger dynamics \citep[\emph{e.g.}][]{ada17} and radial profiles of the ICM thermodynamic properties without relying on X-ray spectroscopic data \citep[\emph{e.g.}][]{rup17}. Furthermore, the comparison between the ICM morphological states derived from X-ray and SZ observations can uncover evidence of ICM disturbances that neither probe is able to highlight independently \citep[\emph{e.g.}][]{ada14,rup18}. For the first time we can characterize all the ICM thermodynamic properties and the morphology of individual clusters from joint X-ray/SZ analyses at $z>1$.\\
\indent The current SZ catalogs do not allow us to define a mass-limited sample of $z>1$ clusters to be considered for high angular resolution SZ and X-ray follow-ups in similar ranges of angular scales. The SPT clusters ($\delta < -20^{\circ}$) cannot be observed by MUSTANG-2 nor by NIKA2 and the $z>1$ ACT clusters have an average declination of $-2.5^{\circ}$ which is also too low to perform efficient SZ observations with these instruments. Furthermore, the Atacama Large Millimeter/submillimeter Array in the compact array configuration is not adapted to probe simultaneously the inner and outer regions of the ICM in reasonable exposure times \citep[\emph{e.g.}][]{kit16}.\\
\indent The current generation of Infra-Red (IR) selected galaxy-based cluster searches complements the work of past millimeter surveys, exploring large volumes of the Universe to detect the rarest high-mass and high-$z$ clusters. The \emph{Massive and Distant Clusters of WISE Survey} (MaDCoWS) has been designed to detect the most massive galaxy clusters at $z \gtrsim  1$ \citep{gon18}. It offers the largest survey area among current high-redshift cluster searches, using infrared and optical imaging from the Wide-field Infrared Survey Explorer \citep[WISE;][]{wri10} and the Panoramic Survey Telescope and Rapid Response System \citep[PanSTARRS;][]{cha16} to robustly isolate galaxy clusters at $z \gtrsim 1$ over more than 80\% of the extragalactic sky. The combination of high angular resolution SZ observations from NIKA2 or MUSTANG-2 with X-ray data measured in the direction of MaDCoWS clusters offers an ideal opportunity to characterize the ICM evolution at $1 < z < 2$.\\
\indent In this paper, we present a multi-wavelength analysis of the MaDCoWS cluster \moo\ at $z=1.19$ confirmed in SZ by \cite{gon15} using the Combined Array for Research in Millimeter-wave Astronomy (CARMA). These previous SZ observations have shown that \moo\ is the most massive cluster known at $z > 1.15$ with a mass $M_{500} = (6.0 \pm 0.9) \times 10^{14}~\mathrm{M_{\odot}}$. By combining spatially resolved X-ray and SZ observations from \chandra\ and NIKA2 we can, for the first time, estimate the radial profiles of all ICM thermodynamic properties, map their average values along the line of sight, and study the relation between the spatial distribution of these properties and cluster morphology at $z>1$. We describe both the NIKA2 observations and raw data analysis in Sect. \ref{sec:nika2_obs}. We characterize the radio source contamination of the SZ signal in the NIKA2 data in Sect. \ref{sec:radio_pts}. In Sect. \ref{sec:chandra_obs}, we give the details of the \chandra\ observations and the X-ray data reduction. We study the cluster morphology based on our multi-wavelength dataset in Sect. \ref{sec:moo_morphology}. We combine the X-ray and SZ data to estimate the radial profiles of the ICM thermodynamic properties in Sect. \ref{sec:moo_1d}. In Sect. \ref{sec:moo_2d}, maps of the ICM temperature and entropy are obtained from the combination of the \chandra\ and NIKA2 data at the pixel level. We present our perspectives given the results of this study in Sect. \ref{sec:perspective} and give a summary of our work in Sect. \ref{sec:conclusions}. In this paper, we assume a $\mathrm{\Lambda}$CDM cosmology based on the latest results from the \planck\ collaboration \citep{pla18}.

\section{NIKA2 SZ observations}\label{sec:nika2_obs}

This section presents the details of the NIKA2 SZ observations of \moo\ completed in October 2017. We first briefly review the main properties of the thermal SZ effect. We then describe the conditions of the data acquisition as well as the different steps of the raw data analysis. We estimate both the residual noise properties and SZ signal filtering and finally characterize the cluster morphology based on the NIKA2 SZ surface brightness maps.

\subsection{The thermal Sunyaev-Zel'dovich effect}\label{subsec:tSZ_effect}

The thermal SZ effect \citep[tSZ;][]{sun72,sun80} corresponds to a variation of the apparent brightness of the cosmic microwave background (CMB) due to the inverse Compton scattering of CMB photons on energetic electrons within any reservoir of hot plasma along the line of sight:
\begin{equation}
        \frac{\Delta I_{tSZ}}{I_0} = y_{tSZ} \, f(\nu, T_e),
\label{eq:deltaI}
\end{equation}
where $f(\nu, T_e)$ characterizes the frequency dependence of the tSZ spectrum \citep{bir99,car02} and $T_e$ is the electronic temperature of the plasma. The Compton parameter $y_{tSZ}$ gives the amplitude of the spectral distortion in the direction $\hat{n}$. It is expressed as:
\begin{equation}
        y_{tSZ}(\hat{n}) = \frac{\sigma_{\mathrm{T}}}{m_{e} c^2} \int P_{e} \, dl,
        \label{eq:y_compton}
\end{equation}
where $m_{e}$ is the electron mass, $c$ the speed of light, $\sigma_{\mathrm{T}}$ the Thomson scattering cross section, and $P_{e}$ is the electron pressure distribution of the gas. Being almost redshift independent\footnote{In practice, the observation of high redshift sources with the SZ effect is only limited by the instrumental beam dilution.}, the thermal SZ effect allows us to directly measure the ICM pressure distribution up to high redshift. It depends only mildly on the ICM temperature through the relativistic corrections to the tSZ spectrum \citep{ito98,poi98}. More details can be found on the information gain brought by spatially resolved SZ observations to probe ICM astrophysics in the review of \cite{mro19}.

\subsection{NIKA2 observations and data reduction}\label{subsec:nika2_data}

We conducted spatially resolved SZ observations of \moo\ in October 2017 with the NIKA2 camera (OpenTime: 082-17, PI: F. Ruppin) installed at the Institut de Radioastronomie Millim\'etrique (IRAM) 30-m telescope. We have observed this cluster for an effective time of 10.4 hours. The pointing center was chosen to be $(\mathrm{R.A., Dec.})_{\mathrm{J2000}}$ = (11:42:46.6, +15:27:15.0) according to the estimate of the SZ centroid position found by \cite{gon15} using CARMA. We defined the scanning strategy in a similar way as the one presented in \cite{ada15} and \cite{rup17}. A succession of on-the-fly raster scans of $8\times 4~\mathrm{arcmin^2}$ with $10$~arcsec steps between each subscan has been realized in four different directions in equatorial coordinates (J2000) at a scanning speed of $40$~arcsec/s. The cluster has been observed at a mean elevation of $55.4^{\circ}$. The weather conditions at the time of the observations were quite good with a mean zenith opacity of 0.19 at 150~GHz, 0.32 at 260~GHz, and a rather stable atmosphere.\\
\indent We use Uranus as a primary calibrator of the data. Following the baseline calibration procedure described in \cite{per19}, we obtain total calibration uncertainties of 6\% and 8\% at 150 and 260~GHz respectively. These estimates take into account the absolute calibration uncertainty of 5\% on the Uranus flux density expectations reported in \cite{mor10}. The distribution of pointing corrections realized at regular intervals during the observations of \moo\ is characterized by a mean of 1.9~arcsec and a standard deviation of 1.2~arcsec. More details on the instrumental performance of the NIKA2 camera at the time of the observations can be found in \cite{ada18} and \cite{per19}.\\\begin{figure*}[t]
\centering
\includegraphics[height=6.4cm]{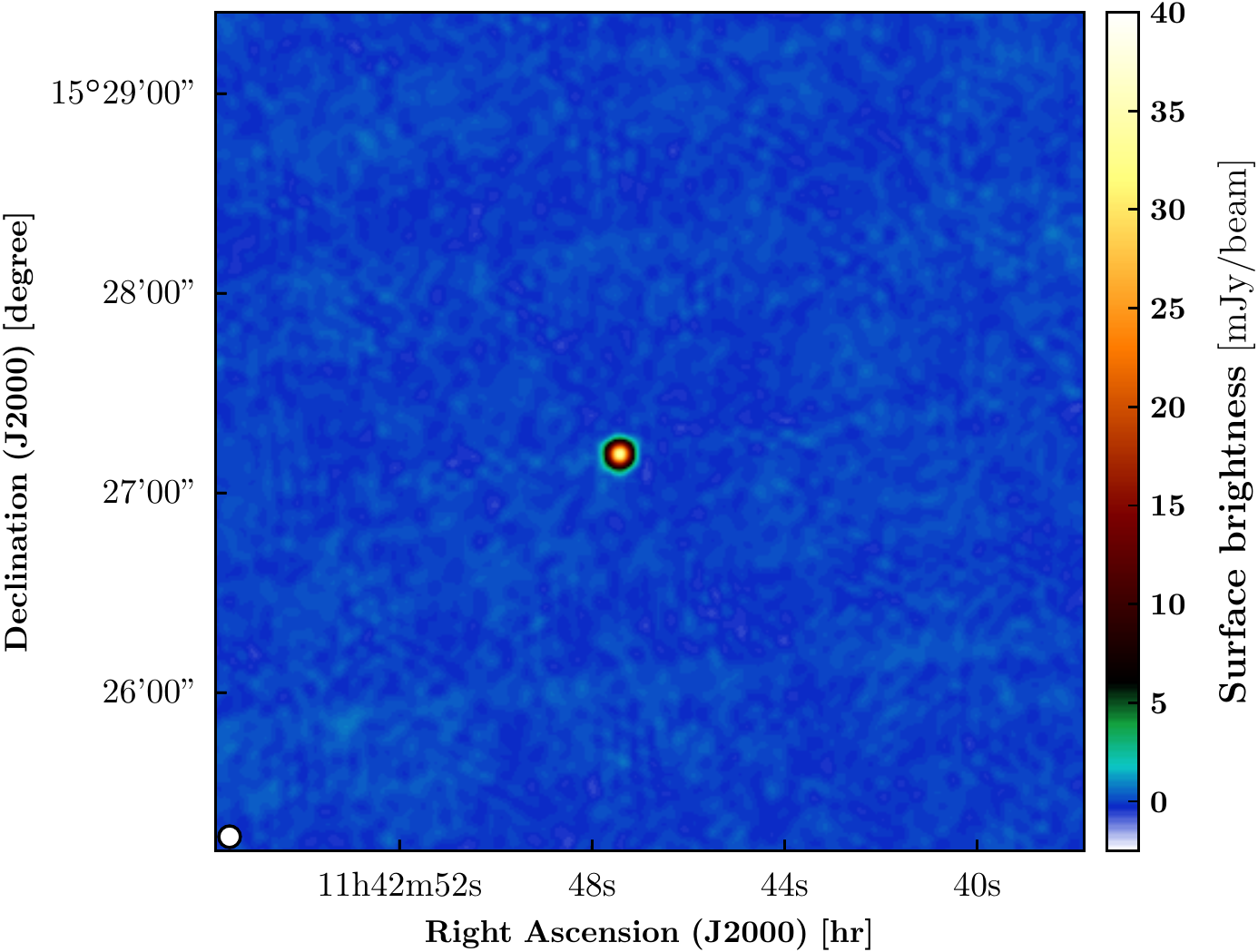}
\hspace{0.6cm}
\includegraphics[height=6.4cm]{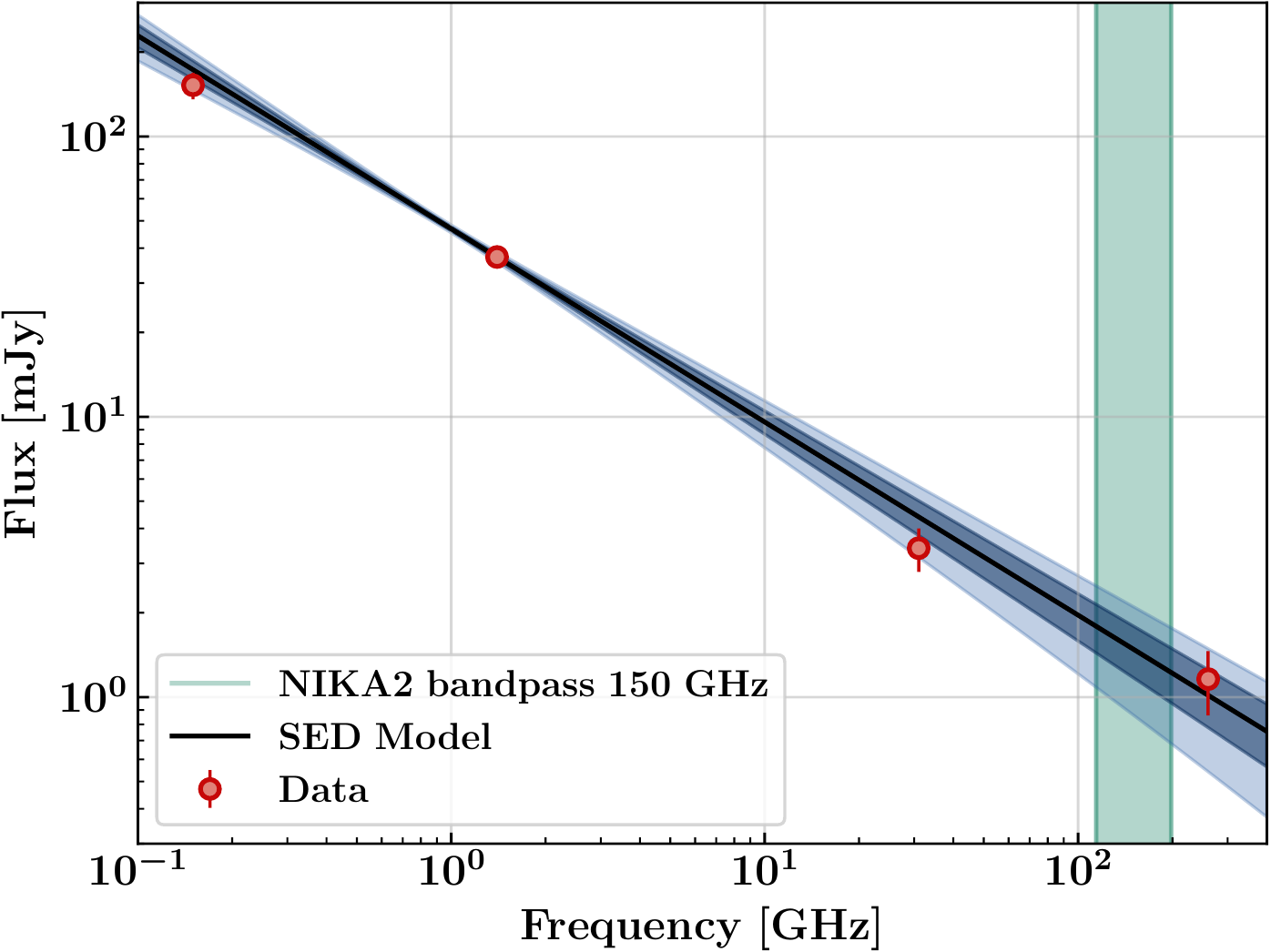}
\caption{{\footnotesize \textbf{Left:} Surface brightness map of the \moo\ sky region obtained by the FIRST survey at 1.4~GHz. A single radio source is detected in the field considered in this paper. \textbf{Right:} Spectral energy distribution (SED) of the radio source shown in the left panel. The red data points correspond to the TGSS, FIRST, CARMA, and NIKA2 260~GHz flux measurements at different frequencies. The black line is our best-fit SED model. The dark and light blue shaded regions give the $1\sigma$ and $2\sigma$ confidence intervals. The green band indicates the range of frequencies covered by NIKA2 at 150 GHz.}}
\label{fig:radio_source}
\end{figure*}\begin{table*}[t]
\begin{center}
\caption{{\footnotesize Location of the radio source detected in a $4$ arcmin radius circular region centered on \moo. We report its flux densities measured at different frequencies by the TGSS \citep{int17} and FIRST \citep{bec95} surveys as well as the CARMA \citep{gon15} and NIKA2 observations of the cluster.}}\label{tab:radio_flux}
\begin{tabular}{ccccc}
\hline
\hline
Position & 153 MHz & 1.4 GHz & 31 GHz & 260 GHz \\
 & [mJy] & [mJy] & [mJy] & [mJy] \\
\hline
11h42m47.421s +15d27m11.48s  & $152.3 \pm 16.6$ & $37.17 \pm 0.20$ & $3.4 \pm 0.6$ & $1.2 \pm 0.3$ \\
\hline
\hline
\end{tabular}
\end{center}
\end{table*}The selection of valid detectors and the removal of cryogenic vibrations and cosmic ray glitches from the raw data have been realized following the pre-processing method detailed in \cite{ada15}. We removed the spatially correlated noise contaminants induced by both the atmospheric emission and the electronic readout system using an iterative procedure similar to the one described in \cite{rup18}. The data measured by each array are treated separately. For each detector timeline, a contaminant template is defined as a combination of a common mode computed using all the valid detectors in the considered array, a common mode estimated across the timelines of detectors sharing the same readout electronic board, and the elevation path of the detector in the plane of the sky. This allows us to simultaneously model the atmospheric and electronic contaminants as well as the timeline drift induced by air mass variations during the scan. At the end of each iteration, we locate the map pixels with a signal-to-noise ratio higher than 4 and remove the corresponding signal amplitude in each detector timeline before the common mode estimations of the next iteration. This allows us to reduce the bias induced by the SZ signal on the estimate of the contaminant templates. The filtering of the SZ signal is therefore reduced at each iteration. We stop the iterative procedure when the variation of the SZ peak amplitude caused by this decrease of the signal filtering is lower than 0.1\%. This corresponds to a total of 13 iterations of the raw data analysis. We project the processed timelines on two different pixelized grids. The first one is defined by a pixel size of 0.984~arcsec equal to the one considered in the \chandra\ analysis described in Sect. \ref{sec:chandra_obs}. This allows us to combine directly the NIKA2 and \chandra\ maps in Sect. \ref{sec:moo_2d} in order to estimate the projected distributions of the cluster thermodynamic properties. For the second one, we use a pixel size of 3~arcsec in order to increase the computing efficiency of the SZ deprojection procedure described in Sect. \ref{sec:moo_1d} without degrading significantly the NIKA2 angular resolution at 150 and 260~GHz. All the individual scans are finally coadded using an inverse variance noise weighting of the data samples in each pixel to obtain the final maps shown in Fig.~\ref{fig:nk2_maps}.\\
\indent The methodology applied to estimate the contaminant templates based on two different common modes allows us to significantly reduce the amount of residual correlated noise in the final maps. The residual noise power specrum has been estimated based on null maps computed from the semi-difference of coadded scans issued from two equivalent subsamples following the procedure described in \cite{ada16} and \cite{rup18}. The noise power spectra measured at 150 and 260~GHz are fitted by a model defined as the sum of a white noise component and a power law to account for the spatial correlations of the residual noise. Although the power law model obtained at 260~GHz significantly differs from a constant, we find that the noise power spectrum measured at 150~GHz is well modeled by a simple white noise component. For this reason, the diagonal elements of the noise correlation matrix at 150~GHz dominate over non-diagonal ones. We can therefore assume that the error associated with the signal measured in the map pixels at 150~GHz is given by the root mean square (RMS) value in each of them. This results in a significant decrease of the computation time of the deprojection procedure detailed in Sect. \ref{sec:moo_1d}.\\
\indent This significant improvement in the correlated contaminant removal comes at the cost of a high filtering of the extended SZ signal at signal-to-noise ratios lower than the $4\sigma$ threshold used in the iterative analysis of the raw data. We compute the circular transfer function resulting from the NIKA2 observations and data processing at 150~GHz using simulations as described in \cite{ada15}. We find that the SZ signal filtering is on average 17\% larger than the one resulting from the analysis procedure described in \cite{rup18} in the range of angular scales that can be constrained by NIKA2, \emph{i.e.}~$[0.25 - 6.5]$~arcmin. However, the SZ map obtained at 150~GHz in \cite{rup18} was significantly contaminated by spatially correlated residual noise. In this paper, we choose to favor an analysis method that allows us to measure simultaneously a nearly flat noise power spectrum at 150~GHz and an extension of the region where the SZ signal is significant that is already larger than the one where the X-ray signal measured by \chandra\ is significantly different from the background. This allows us to maximize the signal-to-noise ratio in the maps of the ICM temperature and entropy obtained in Sect. \ref{sec:moo_2d}.

\subsection{NIKA2 maps of \moo}\label{subsec:moo_sz}

The surface brightness maps of \moo\ obtained at the end of the analysis of the NIKA2 data at 150 and 260~GHz are shown in Fig.~\ref{fig:nk2_maps}. They correspond to the grids made of 0.984~arcsec pixels and centered on the SZ peak coordinates measured by CARMA \citep{gon15}. For visual purposes, we have smoothed the maps with an additional 8 and 5~arcsec FWHM Gaussian filter at 150 and 260~GHz respectively. We measure the RMS noise at the map center to be $98~\mathrm{\mu Jy/beam}$ and $423~\mathrm{\mu Jy/beam}$ at these effective resolutions of 20 and 13~arcsec at 150 and 260~GHz respectively. These estimates have been obtained following the procedure described in \cite{ada17}. Simulated noise maps are computed from the NIKA2 residual noise power spectra at 150 and 260~GHz and from astrophysical contaminant models (\citealt{pla14}, \citealt{bet12}, and \citealt{tuc11}). The latters allow us to take into account the contributions induced by both the cosmic infrared background (CIB) shot noise and clustering and the CMB temperature anisotropies in the residual noise. As the processing methodology adopted to analyze the NIKA2 raw data has enabled reducing the amount of residual correlated noise compared to previous NIKA2 studies, the instrumental and atmospheric noise contributions are comparable to the one induced by CIB shot noise at 150~GHz. We add all the noise contaminants in quadrature in order to estimate the signal-to-noise ratio in each map pixel at 150 and 260~GHz.\\

We observe significant negative SZ signal in the NIKA2 map at 150~GHz shown in the left panel of Fig.~\ref{fig:nk2_maps}. The SZ peak is found at a position of $(\mathrm{R.A., Dec.})_{\mathrm{J2000}}$ = (11:42:45.9, +15:27:12.0) with a significance of $17\sigma$ at ${\sim}10$~arcsec from the pointing center. We compute the SZ surface brightness profile from the NIKA2 map at 150~GHz and measure a signal-to-noise ratio (S/N) higher than $3$ up to $1.1$~arcmin away from the SZ peak. This is comparable to the extension of the diffuse X-ray emission recovered by \chandra\ as described in Sect. \ref{sec:chandra_obs}. We find some evidence of an elliptical morphology of the SZ signal with an E-W orientation. We fit the $3\sigma$ S/N contour with an ellipse and measure a flattening $f = 1 - b/a = 0.19$ where $a$ and $b$ are the major and minor axis length respectively. This is consistent with the SZ morphology found by CARMA \citep{gon15}. However, the NIKA2 instrumental performance allows us to further resolve the ICM structure. This will enable constraining the radial pressure profile in Sect. \ref{sec:moo_1d}.\\
\indent We do not detect any significant SZ signal in the NIKA2 260~GHz map shown in the right panel of Fig.~\ref{fig:nk2_maps} as anticipated given the expected SZ signal and the noise level at this frequency. Based on the value of the peak SZ surface brightness at 150~GHz, the tSZ spectrum analytic expression, and the NIKA2 bandpasses at the time of the observation, the expected value of the peak SZ surface brightness at 260~GHz is found to be $540~\mathrm{\mu Jy/beam}$. This corresponds to 1.3 times the RMS noise found at the NIKA2 map center at this frequency. However, we detect several point sources in the NIKA2 map at 260~GHz including one within the region where significant SZ signal is measured at 150~GHz. This source is detected at $25$~arcsec to the east of the NIKA2 SZ peak at 150~GHz. This position is consistent with a radio source found in the Faint Images of the Radio Sky at Twenty-Centimeters survey \citep[FIRST;][]{bec95}, which covers the northern sky at 1.4~GHz. We highlight this position with a green cross in both NIKA2 maps in Fig.~\ref{fig:nk2_maps}. The signal emitted by this source is partly compensating the negative SZ signal induced by \moo\ at 150~GHz. This explains the origin of the hole in the SZ signal found at this position in the NIKA2 map. For this reason, it is essential to estimate the expected flux of this source at 150~GHz before constraining the ICM pressure profile from a deprojection of the NIKA2 data to minimize the bias induced by this contaminant.

\begin{figure*}[t]
\centering
\includegraphics[height=6.3cm]{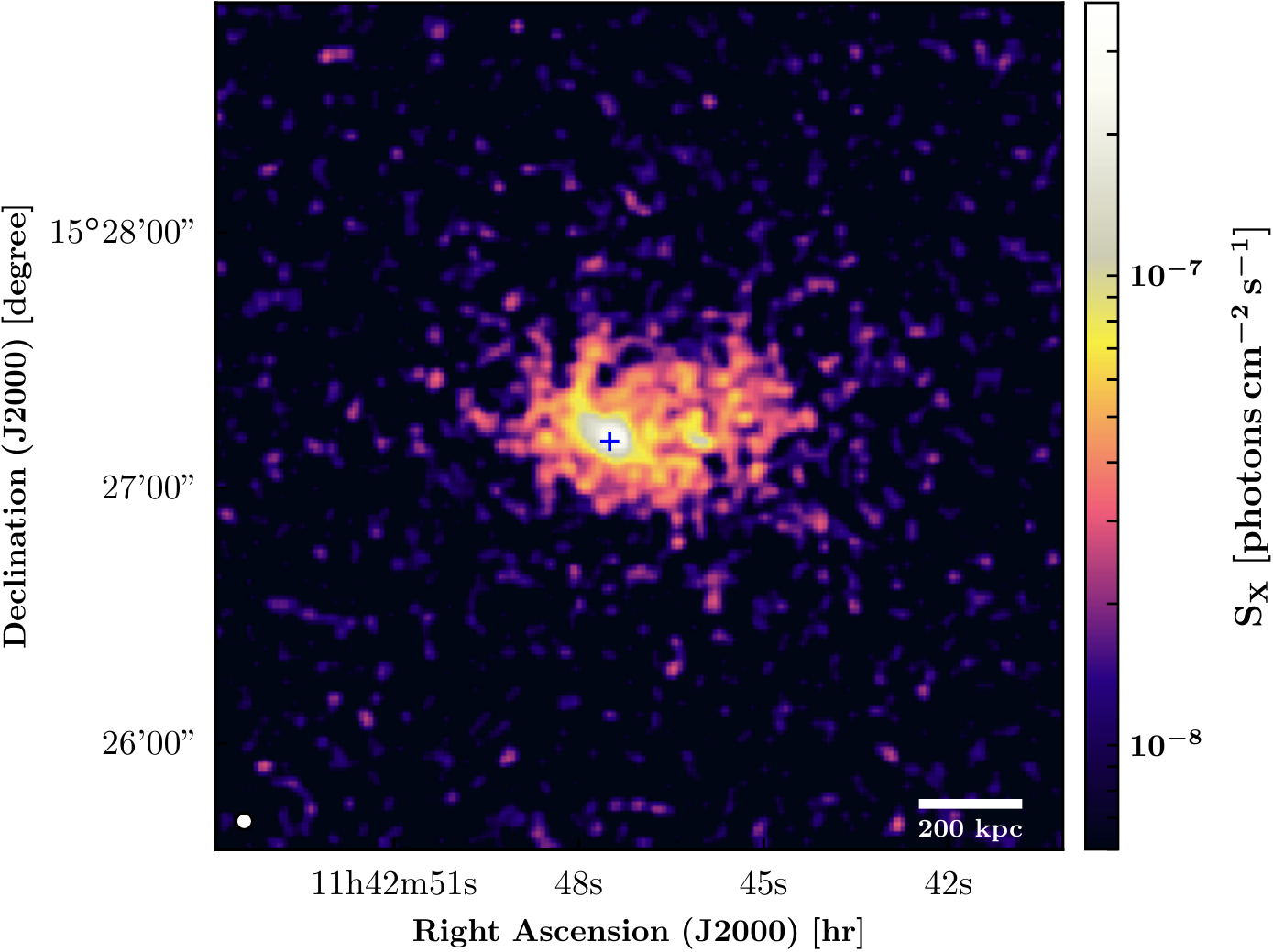}
\hspace{0.6cm}
\includegraphics[height=6.3cm]{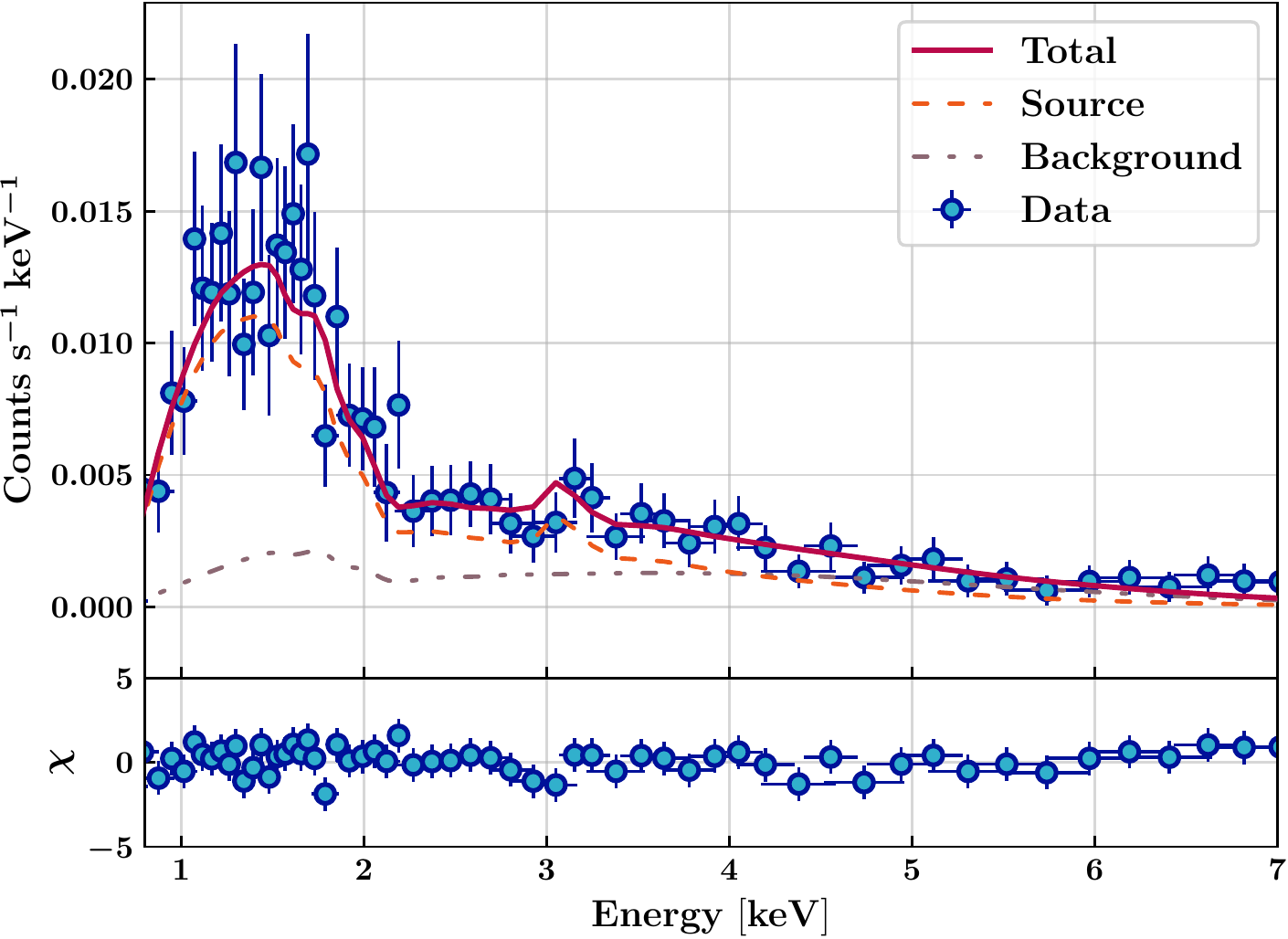}
\caption{{\footnotesize \textbf{Left:} Exposure-corrected \chandra\ flux map of \moo\ in the 0.7-7.0~keV band. We have binned the event list using 0.984~arcsec pixels. The image has been smoothed with an additional 3~arcsec FWHM Gaussian filter and the colors are displayed with a logarithmic scaling for display purposes. The blue cross gives the location of the radio source characterized in Sect. \ref{sec:radio_pts}. \textbf{Right:} X-ray spectrum extracted from the \chandra\ event list after subtracting the particle background (blue points) in an annulus centered on the cluster large scale centroid with inner and outer radii of $0.15R_{500}$ and $R_{500}$ respectively. The sum of the source model (orange) and the astrophysical background model (mauve) fitted jointly gives the final model (red). The lower panel shows the ratio between the difference of the data and the model with the uncertainty associated with each data point.}}
\label{fig:chandra_raw}
\end{figure*}

\section{Point source contamination}\label{sec:radio_pts}

This section is dedicated to the study of the point source contamination of the SZ signal measured in the NIKA2 map at 150~GHz (see Sect. \ref{subsec:moo_sz}). This work will allow us to fit jointly the SZ signal and the point source emission in order to accurately estimate the ICM pressure profile in Sect. \ref{sec:moo_1d} \citep[\emph{e.g.}][]{say13}.\\
As shown in the left panel of Fig.~\ref{fig:radio_source}, a single radio source has been detected by the FIRST survey at 1.4~GHz in the region observed by NIKA2 \citep{bec95}. This source is located on the east side of the SZ peak of \moo\ at the same location of the X-ray peak emission measured by \chandra\ (see Sect. \ref{sec:chandra_obs}). Its location also coincides with that of the brightest cluster galaxy (BCG) observed in infrared by \emph{Spitzer} (see Sect. \ref{sec:moo_morphology}). We therefore conclude that this radio source is hosted by the BCG. The flux of this source has also been measured at 153~MHz by the TIFR GMRT Sky Survey \citep[TGSS;][]{int17} and at 31~GHz by CARMA \citep{gon15}. We measure the flux of this source in the NIKA2 map at 260~GHz by fitting a 2D Gaussian function at the source location found by FIRST using a FWHM fixed to the NIKA2 angular resolution at this frequency. We report our result along with the previous flux measurements of this source in Table \ref{tab:radio_flux}. Following the methodology developed in \cite{ada16}, we estimate the expected flux of this radio source at 150~GHz by fitting its spectral energy distribution (SED) based on the available flux measurements. We use a power law model given by:
\begin{equation}
F_{\nu} = F_{\rm{1\,GHz}}\left(\frac{\nu}{\rm{1~GHz}}\right)^{\alpha_{\rm{radio}}}
\label{eq:p_law_sed}
\end{equation}
where $F_{\rm{1\,GHz}}$ gives the SED amplitude at a reference frequency of 1~GHz and $\alpha_{\rm{radio}}$ is the SED spectral index. The four data points shown in red in the right panel of Fig.~\ref{fig:radio_source} are used to obtain the best-fit values of these two parameters, \emph{i.e.}~$(F_{\rm{1\,GHz}},\alpha_{\rm{radio}}) = (46.8~\rm{mJy},-0.69)$, and their associated covariance at maximum likelihood. We simulate mock SEDs using a Gaussian sampling of the parameter space around the best-fit values based on the parameter covariance. The best-fit SED model is shown with the black line in Fig.~\ref{fig:radio_source} along with its associated $1\sigma$ and $2\sigma$ confidence intervals in dark and light blue obtained from the simulated SEDs. Each mock SED is integrated in the NIKA2 bandpass at 150~GHz. The mean and standard deviation of all the realizations give us an estimate of the expected flux of the radio source at this frequency, $F_{\rm{150~GHz}} = 1.5 \pm 0.3~\rm{mJy}$. We use this result to define a Gaussian prior on the source emission at 150~GHz in Sect \ref{sec:moo_1d}.\\

\begin{figure*}[t]
\centering
\includegraphics[height=6.4cm]{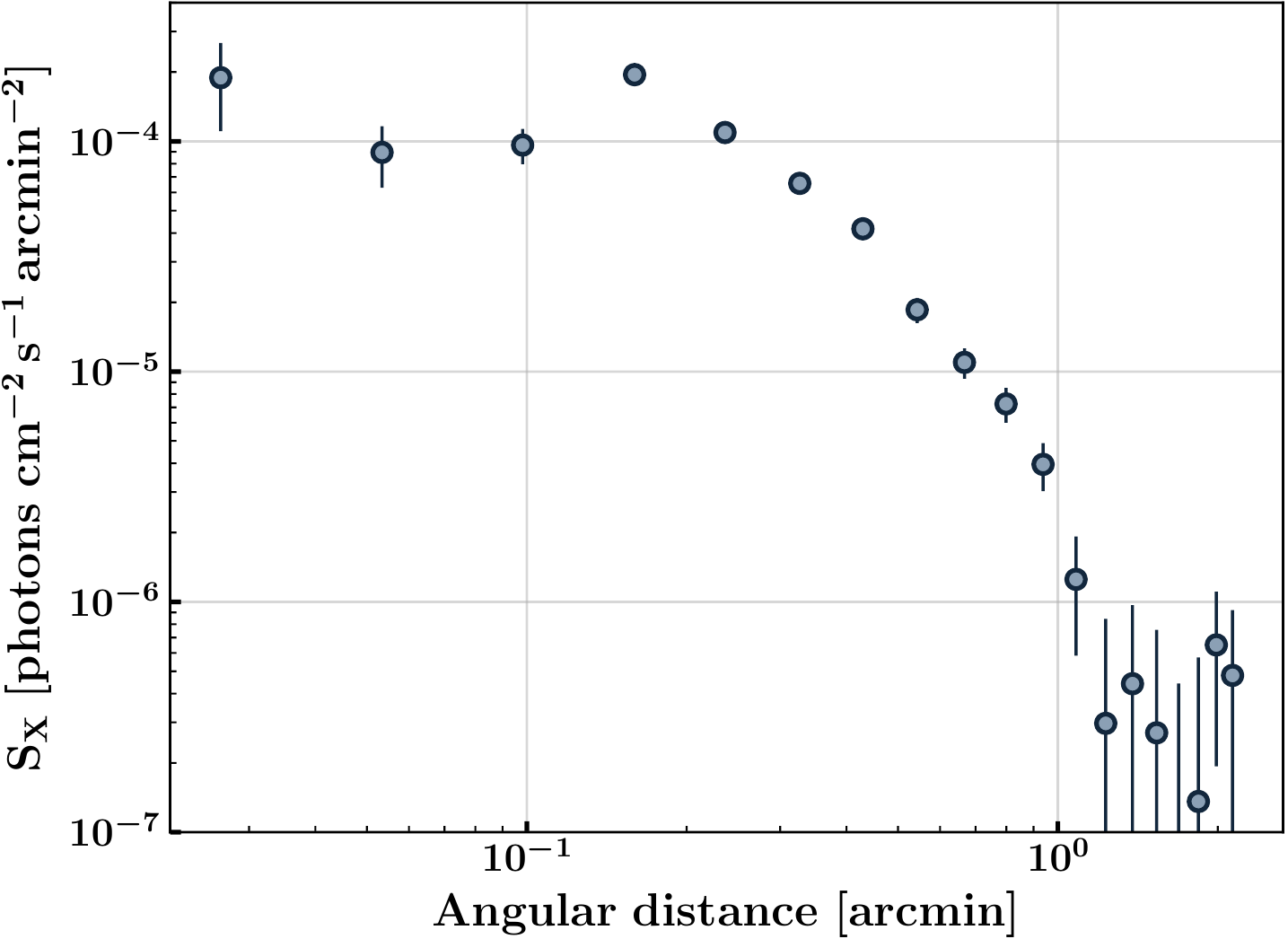}
\hspace{0.6cm}
\includegraphics[height=6.4cm]{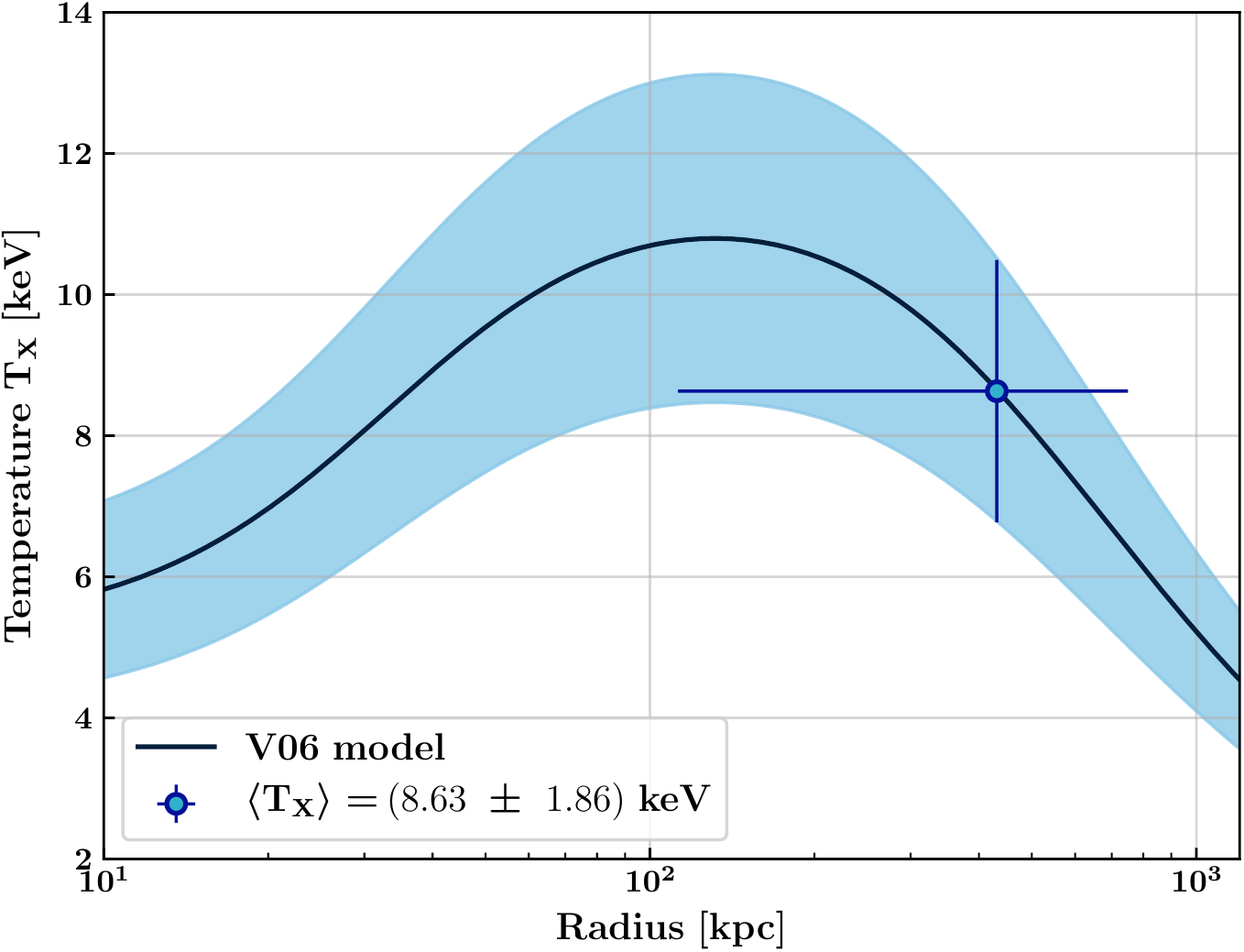}
\caption{{\footnotesize \textbf{Left:} \chandra\ X-ray surface brightness profile extracted within the 20 annuli defined by Eq. \ref{eq:XSB_an} centered on the cluster large scale centroid. \textbf{Right:} Mean ICM spectroscopic temperature estimated from the analysis of the X-ray spectrum measured by \chandra\ between $0.15R_{500}$ and $R_{500}$ (blue point). The universal temperature model defined by \cite{vik06} scaled to our measurement is shown with a black line and the blue region gives the $1\sigma$ confidence interval.}}
\label{fig:chandra_1stprod}
\end{figure*}

There is no available  ancillary data on the contamination induced by submillimeter galaxies in the considered sky region. However, based on the NIKA2 data at 260~GHz, we find no submillimeter sources with a flux high enough to be detected within the region where significant SZ signal is measured at 150~GHz. We therefore consider this contaminant to be negligible in the following sections. We note however that a submillimeter galaxy is detected at 260~GHz with a signal-to-noise ratio of 6 at a position of $(\mathrm{R.A., Dec.})_{\mathrm{J2000}}$ = (11:42:50.8, +15:25:50.4), \emph{i.e.}~at ${\sim}2.1$~arcmin from the SZ peak at 150~GHz. We compute the ratio between the fluxes measured at this position at 150 and 260~GHz and find a value of $F_{\rm{150~GHz}}/F_{\rm{260~GHz}} = 0.1$. This is consistent with the ratios found for the population of dusty galaxies detected at 150 and 214~GHz by the South Pole Telescope \citep{vie10}.

\section{Chandra X-ray observations}\label{sec:chandra_obs}

This section presents the X-ray analysis of the \chandra\ observations of \moo\ realized during Cycle 17 in February 2017 (ObsID: 18277, PI: A. Stanford). We first review the main steps of the raw data analysis that aim at obtaining both the background and point source-subtracted event list, and the X-ray spectrum in the region of interest. We further present the analysis of these data products in order to measure the cluster X-ray surface brightness profile and its mean spectroscopic temperature.

\subsection{Observations and data reduction}\label{subsec:chandra_data}

The X-ray observations of \moo\ were obtained in the VFAINT data mode for a total exposure of 46.96~ks using the Advanced CCD Imaging Spectrometer (ACIS) I-chips on board the \chandra\ X-ray Observatory. We follow the data reduction procedure described in \cite{mcd17} and references therein. We have reduced the data using the Chandra Interactive Analysis of Observations (CIAO) software v4.10 based on the calibration database (CALDB) v4.8.0 provided by the \chandra\ X-ray Center (CXC). We have used the \texttt{chandra\_repro} script to reprocess the level 1 event files using the latest charge transfer inefficiency corrections and time-dependent gain adjustments. The ACIS particle background for very faint mode observations is also cleaned based on outer pixel pulse heights during this analysis step. We use the \texttt{lc\_clean} routine based on M. Markevitch's program \citep{mar01} to remove flares from lightcurves created with a temporal bin size of 259.28 seconds. We find a total, cleaned exposure time of 46.19~ks.\\
\indent The exposure map associated with the observations has been computed in an energy band restricted from 0.5~keV to 7~keV, and a center-band energy of 2.3 keV as recommended by the CXC. Point sources have been identified using the \texttt{wavdetect} script based on a wavelet decomposition technique \citep[\emph{e.g.}][]{vik98}. We also perform a visual inspection of the regions enclosing the detected sources. A mask has been generated using the resulting list of point sources. It is used in Sect. \ref{subsec:moo_xray} to produce a cleaned event list from which the X-ray surface brightness profile as well as the X-ray spectrum are extracted. We define the X-ray background as a combination of a particle and instrumental background and a local astrophysical background. For the particle and instrumental background, we normalize unscaled stowed background files to the count rate observed in the 9-12~keV band. The significant diffuse emission from the cluster has only been detected by the I3 chip. The remaining three ACIS-I chips are therefore used as an estimate of the local astrophysical background once both particle background and point sources have been removed.\\

We show an exposure-corrected map of the \chandra\ observations of \moo\ after background and point source subtraction in the left panel of Fig.~\ref{fig:chandra_raw}. The diffuse X-ray emission is significantly detected within a region of similar angular scale as the one obtained with the NIKA2 SZ observations (see Fig.~\ref{fig:nk2_maps}). The X-ray peak (blue cross) is detected at the same location as the radio source observed in the FIRST map in the left panel of Fig.~\ref{fig:radio_source}. We note that a careful analysis has been made in order to show that the X-ray peak is not contaminated by an X-ray point source but is really caused by an over-density within the ICM. In particular, the \texttt{wavdetect} routine has been used in different energy intervals within the soft and hard bands to confirm that no X-ray point source is detected at the X-ray peak position. Therefore, together with the results obtained on the radio emission of the central AGN in Sect. \ref{sec:radio_pts}, we find that the latter has an X-ray luminosity compatible with 0 and a radio luminosity at 1.4~GHz of $L_{1.4 \, \mathrm{GHz}} = (1.373 \pm 0.007) \times 10^{25}~\mathrm{W.Hz^{-1}}$. This clearly indicates that the BCG hosting the AGN is a member of the corona class and should have an X-ray cool core according to \cite{sun09}. The cluster emission is clearly elongated westward with respect to the X-ray peak. For this reason, the large-scale morphology of the cluster is similar to the one observed in the NIKA2 map at 150~GHz (see Sect. \ref{subsec:moo_sz}).\\
\indent As the X-ray emission is apparently not azimuthally symmetric, we choose to use both the cluster large-scale centroid and the X-ray peak for all estimates shown in this paper unless otherwise noted. The centroid position is computed iteratively within a circular region of 1~arcmin radius centered on the last estimate of the centroid location. We initialize this position to the X-ray peak one. A comparison between the locations of the X-ray peak, the X-ray centroid, and the SZ peak is discussed in Sect. \ref{sec:moo_morphology}.\\
\indent Following \cite{mcd16}, we extract the X-ray spectrum in the 0.7-7.0~keV band using the \texttt{specextract} script in a core-excised circular annulus centered on the large-scale centroid and mapping the radii $0.15 R_{500}<r<R_{500}$. We iteratively estimate the value of $R_{500}$ based on the best-fit value of the spectroscopic temperature $T_X$ (see Sect. \ref{subsec:moo_xray}) and the $M_{500}{-}T_X$ scaling relation from \cite{vik09} and find a value of $790\pm 87$~kpc. The scaled stowed background files are used to compute the spectrum of the particle background to be subtracted from the measured spectrum. We repeat the same procedure in the remaining three ACIS-I chips to measure the spectrum of the local astrophysical background. The final X-ray spectrum measured in the cluster region is shown with blue points in the right panel of Fig.~\ref{fig:chandra_raw}. We group the energy channels to obtain a signal-to-noise ratio higher than 5 in each bin. This spectrum presents enough statistics to enable estimating the mean spectroscopic temperature of the ICM.

\begin{figure*}[t]
\centering
\includegraphics[height=6.9cm]{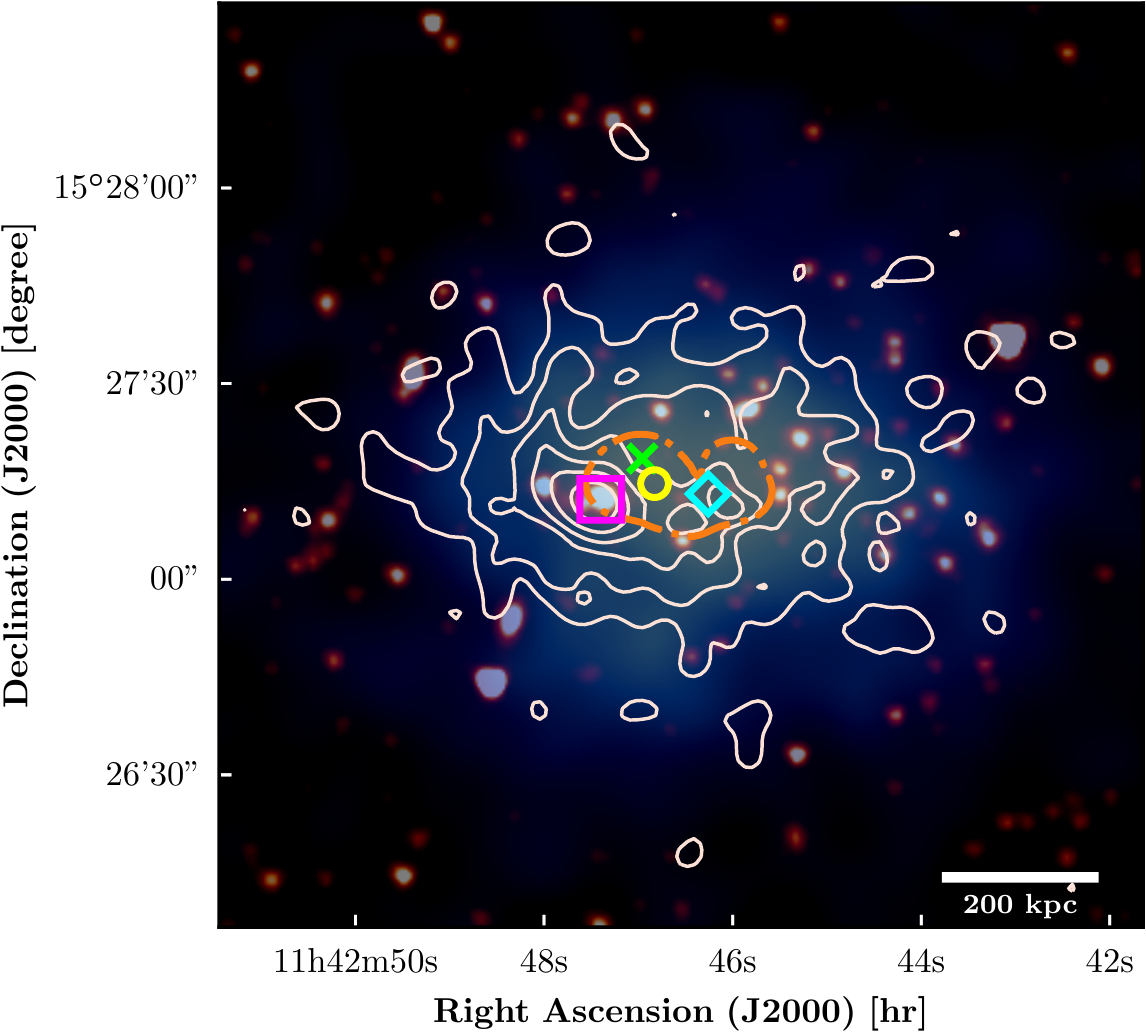}
\hspace{0.1cm}
\includegraphics[height=6.9cm]{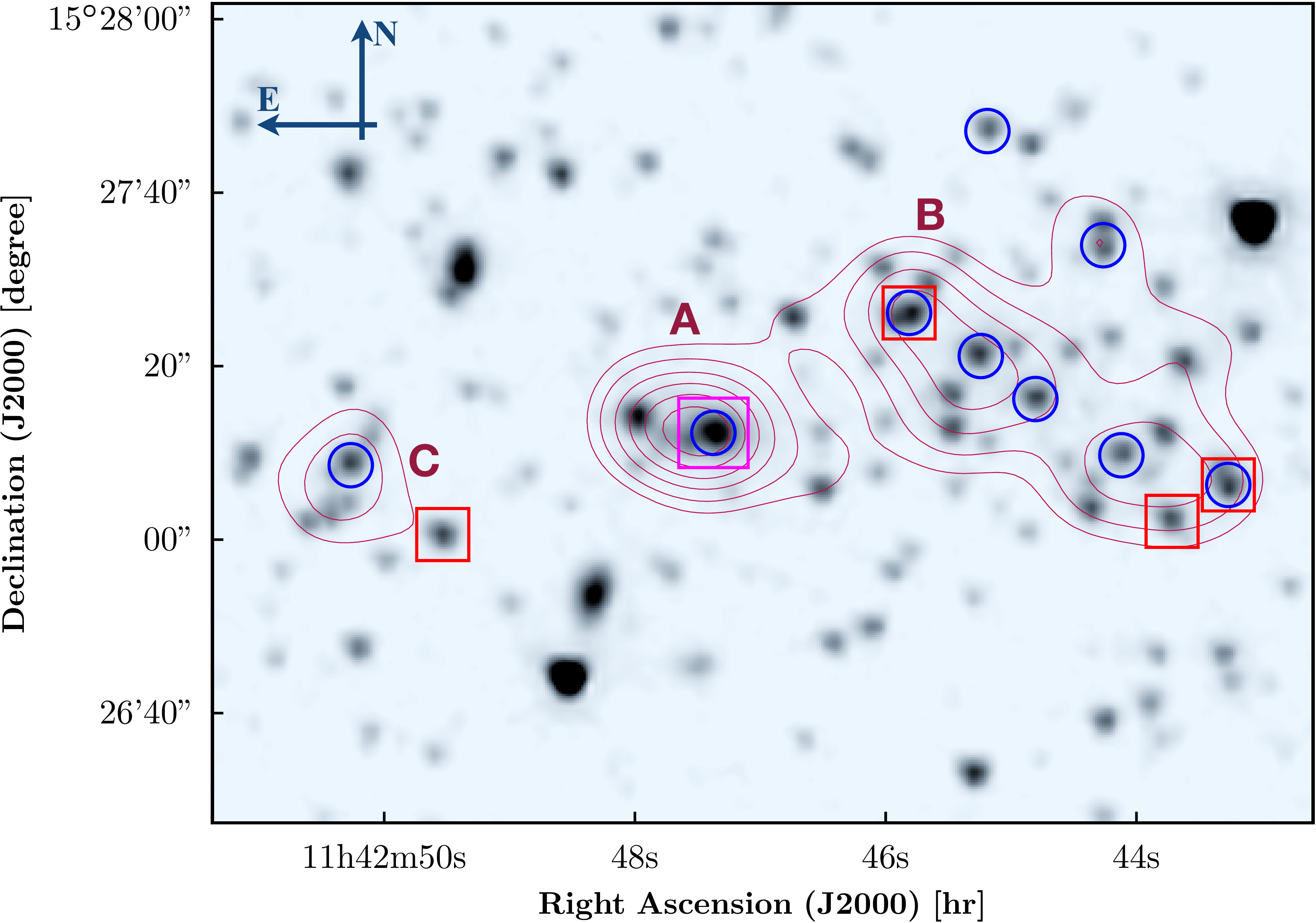}
\caption{{\footnotesize \textbf{Left:} Multi-wavelength map of \moo\ showing the \chandra\ X-ray count isocontours (white), the NIKA2 SZ signal measured at 150~GHz (blue), and the galaxy distribution from IRAC [3.6] (red). The SZ peak location before the point source subtraction is shown with the cyan diamond. The SZ peak location after subtracting the source signal contamination is shown with the green cross. The closed area delimited by an orange dash-dotted line gives the 95\% confidence region where the SZ peak is found given the uncertainty on the radio source flux and the residual noise fluctuations in the NIKA2 map. The yellow circle gives the position of the large-scale X-ray centroid measured by \chandra. \textbf{Right:} IRAC [3.6] galaxy distribution in the \moo\ field. Red contours give the density peaks in the galaxy distribution. The main peaks are labeled A, B, and C. Red squares give the location of spectroscopic cluster members and blue circles show the locations of individual WISE sources identified as candidate $z > 0.75$ galaxies. In both panels, the magenta square gives simultaneously the location of the \chandra\ X-ray peak, the FIRST radio source, and the IRAC brightest cluster galaxy.}}
\label{fig:rgb_map}
\end{figure*}

\subsection{Mean spectroscopic temperature and surface brightness profile}\label{subsec:moo_xray}

With low S/N X-ray data, it is not possible to extract a temperature profile. Hence, we evaluate a mean spectroscopic temperature from the fitting of the cluster spectrum shown in the right panel of Fig.~\ref{fig:chandra_raw}. The cluster spectrum is fitted jointly with the astrophysical background spectrum using CIAO's \emph{Sherpa} package. We fit the whole spectrum using a combination of \texttt{XSPEC} models \citep[v12.10.0e;][]{arn96} including a single-temperature plasma \citep[\texttt{APEC};][]{smi01} for the cluster emission, a soft X-ray Galactic background (\texttt{APEC}, $k_BT_X = 0.18~\mathrm{keV}$, $Z = Z_{\odot}$, $z=0$) combined with a hard X-ray cosmic spectrum (\texttt{BREMSS}, $k_BT_X = 40~\mathrm{keV}$) for the astrophysical background, and a Galactic absorption model (\texttt{PHABS}). For the latter, we fix the Galactic column density to the value found by \cite{kal05} at this latitude: $n_H = 2.92\times 10^{20}~\mathrm{cm^{-2}}$. We also fix the cluster  redshift to the value estimated from the combination of the Gemini/GMOS, Keck/DEIMOS, and Keck/MOSFIRE spectroscopy measurements presented by \cite{gon15}, \emph{i.e.}~$z=1.19$, in the cluster emission model. We allow the ICM spectroscopic temperature and the different model normalizations to vary in the analysis. Given the cluster redshift, an iron emission line is expected at an energy of ${\sim}3$~keV. As it is not significantly detected in the measured spectrum, we choose to fix the ICM metallicity to $Z = 0.3 Z_{\odot}$.\\
The best-fit total model is shown with the red line in the right panel of Fig.~\ref{fig:chandra_raw}. It is given by the sum of the cluster emission model (orange) and the astrophysical background model (grey). The lower panel of the figure displays the ratio of the difference between the data and the best-fit model with the measurement uncertainty in each energy bin. We do not measure any deviation larger than $3\sigma$ and no significant systematic effect is identified in these residuals. We measure a reduced $\chi^2$ of $1.15 \pm 0.18$ for this analysis based on 59 degrees of freedom. We find that the cluster spectroscopic temperature is given by $T_X = 8.63 \pm 1.86~\mathrm{keV}$ within a radius range $119 < r < 790~\mathrm{kpc}$. This single temperature measurement is shown as a blue point in the right panel of Fig.~\ref{fig:chandra_1stprod}. It is a key parameter to define the X-ray emission measure profile of the cluster in Sect. \ref{sec:moo_1d}. As has been shown by \cite{bou08}, the variation of the neutral hydrogen column density across the field of view can have a significant impact on the ICM temperature estimate. Therefore, a second analysis has been done with a free Galactic column density value. The best-fit value of this parameter is found to be $n_H = 14.68\times 10^{20}~\mathrm{cm^{-2}}$ with an associated ICM temperature of $T_X = 7.75 \pm 2.13~\mathrm{keV}$. As the fitted hydrogen column density is much higher than the expected one, the normalization of the cluster model increases to compensate the absorption that is significant between 0.7 and ${\sim}2$~keV. This explains why we find a lower spectroscopic temperature associated with this increased Galactic absorption. We find a reduced $\chi^2$ of $1.14 \pm 0.18$ for this second analysis based on 58 degrees of freedom. Setting the Galactic column density as a free parameter does not significantly improve the spectrum fit. Furthermore, the temperature estimates obtained with the two analyses are compatible and the fitted value of $n_H$ is in tension with the expected one. Therefore, we choose to keep our first temperature estimate obtained with a fixed hydrogen column density in the following.\\

Following the methodology introduced by \cite{mcd17}, we extract the cluster surface brightness profile in the 0.7-2.0~keV band in 20 annuli defined by:
\begin{equation}
r_{\mathrm{out},i} = (a+bi+ci^2+di^3)R_{500}~~i=1...20
\label{eq:XSB_an}
\end{equation}
where $(a,b,c,d) = (13.779, −8.8148, 7.2829, −0.15633) \times 10^{-3}$. This definition of the radial binning has been optimized to enable an efficient sampling of the X-ray surface brightness profile of galaxy clusters from \chandra\ observations up to $z{\sim}2$. We use the \texttt{dmextract} routine to extract a surface brightness profile from the event list masking the identified point sources. The profile is estimated using both the X-ray peak and the large-scale centroid. We correct the surface brightness profiles for the vignetting effect based on the normalized exposure map estimated in the same energy band. The estimated profile centered on the large-scale centroid  is shown in the left panel of Fig.~\ref{fig:chandra_1stprod}. The inner slope of the profile is relatively constant, which is expected because the large-scale centroid is offset westward with respect to the X-ray peak location (see Sect. \ref{sec:moo_morphology}). Significant constraints are established up to an angular distance from the centroid of ${\sim}1.1~\mathrm{arcmin}$ which corresponds to a physical distance of $562~\mathrm{kpc}$ at the cluster redshift. As this detection radius is similar to the one found with the NIKA2 SZ data at 150~GHz, we will highlight it in each figure of the estimated ICM profiles obtained in Sect. \ref{sec:moo_1d}. The slight increase in surface brightness observed at an angular distance of ${\sim}0.15~\mathrm{arcmin}$ is due to the X-ray peak brightness averaged in the fourth annuli defined by Eq. (\ref{eq:XSB_an}). This \chandra\ X-ray surface brightness profile along with the spectroscopic temperature estimate are the key inputs that are considered in Sect. \ref{sec:moo_1d} to deproject the ICM density profile.

\section{Multi-wavelength analysis: ICM morphology}\label{sec:moo_morphology}

This section aims at describing the morphological properties of \moo\ from a multi-wavelength comparison of the available data-sets. We also infer a possible scenario of the cluster dynamics that would explain the observed cluster morphology in X-ray, optical/infrared, and SZ.\\

We show a multi-wavelength map of \moo\ in the left panel of Fig.~\ref{fig:rgb_map}. It combines the NIKA2 SZ signal with the point source subtracted (blue), the galaxy distribution observed at $3.6~\mu\mathrm{m}$ by the Infrared Array Camera (IRAC) on board \emph{Spitzer} (red), and the \chandra\ X-ray count isocontours (white). The X-ray peak emission is located at the same position as the radio source shown in Fig.~\ref{fig:radio_source} and is marked by a magenta square. The X-ray large scale centroid (see Sect. \ref{subsec:chandra_data}) is found at a distance of ${\sim}100~\mathrm{kpc}$ westward (yellow circle). The angular separation between these two positions is a clear indication of morphological disturbance.\\
The location of the SZ peak found by NIKA2 at 150~GHz depends strongly on the estimate of the radio source flux considered to obtain the point source subtracted map and on the residual noise fluctuations at the map center. For this reason, we have generated 1000 realizations of cleaned SZ maps based on the NIKA2 map shown in the left panel of Fig.~\ref{fig:nk2_maps} and different estimates of the radio source flux given the constraints obtained in Sect. \ref{sec:radio_pts}. We have also added a noise map realization to each cleaned SZ map based on the RMS noise measured at 150~GHz (see Sect. \ref{subsec:moo_sz}). These simulated SZ maps are therefore characterized by an RMS noise increased by a factor $\sqrt{2}$ with respect to the NIKA2 SZ map at 150~GHz shown in the left panel of Fig.~\ref{fig:nk2_maps}. The closed region delimited by an orange dash-dotted line in Fig.~\ref{fig:rgb_map} contains all locations where the SZ peak is found for at least 95\% of all realizations. It contains both the X-ray peak emission and large-scale centroid. The SZ peak found in the NIKA2 raw map at 150~GHz is shown by a cyan diamond. The one measured with the point source subtracted map considering a radio source flux of 1.5~mJy is given by the green cross.\\
The conclusion of this study is twofold. First, we show that the highest ICM thermal pressure is reached within a region that extends over ${\sim}230~\mathrm{kpc}$ along the R.A. axis. This is a clear sign of dynamical disturbance in the cluster core. Second, the thermal pressure excess is almost always found to the west of the X-ray peak. This may imply that the maximum thermal pressure value is not due to a maximum ICM electron density but to a gas temperature excess at this location.\\
We show contours of the IRAC galaxy density distribution observed in this field in the right panel of Fig.~\ref{fig:rgb_map} (red lines). They have been obtained by smoothing the IRAC image with a 10~arcsec FWHM Gaussian kernel, masking the brightest foreground sources. Sources identified as candidate $z > 0.75$ galaxies from the WISE color, magnitude, and quality cuts are highlighted with the blue circles and spectroscopically confirmed cluster members are shown with the red and magenta squares \citep{gon15}. The main galaxy density peak is labeled A. It coincides with the location of the \chandra\ X-ray peak and it contains the BCG (magenta square). A second peak in the spatial distribution of cluster galaxies is found to the west of the thermal pressure excess estimated from the NIKA2 data. It is labeled B in this figure. We note that a small group of galaxies, labeled C, is found to the east of the X-ray peak but is not associated with any X-ray or SZ signal in the \chandra\ and NIKA2 map at 150~GHz. The combination of all the features found from this multi-wavelength analysis of the cluster morphology leads us to the conclusion that \moo\ is an on-going merger.\\

A possible explanation for all the morphological features observed in Fig.~\ref{fig:rgb_map} would be that an infalling subcluster associated with the second peak in the galaxy distribution (B) is merging through the main halo centered on the BCG (A) from the northwest to the south-east regions of the ICM. Such a merger event would shock-heat the gas and induce both a thermal pressure excess and an elongation of the gas distribution to the west of the main halo core. We favor an on-going merger scenario in contrast with a post-merger one because the second galaxy density peak is observed at the same location as the northwest ICM extension. As galaxies are collisionless and the gas is not, we would measure significant differences between the morphology of the ICM and the galaxy distribution if the infalling galaxy group had already passed through the cluster core \citep{mar04}. Furthermore, we measure a relatively low central entropy and cooling time at the position of the BCG (see Sect. \ref{subsec:T_K_prof}). Thus, the cluster core does not seem to have undergone a major merger yet.

\section{Multi-wavelength analysis: ICM profiles}\label{sec:moo_1d}

\begin{figure*}[t]
\centering
\includegraphics[height=5.8cm]{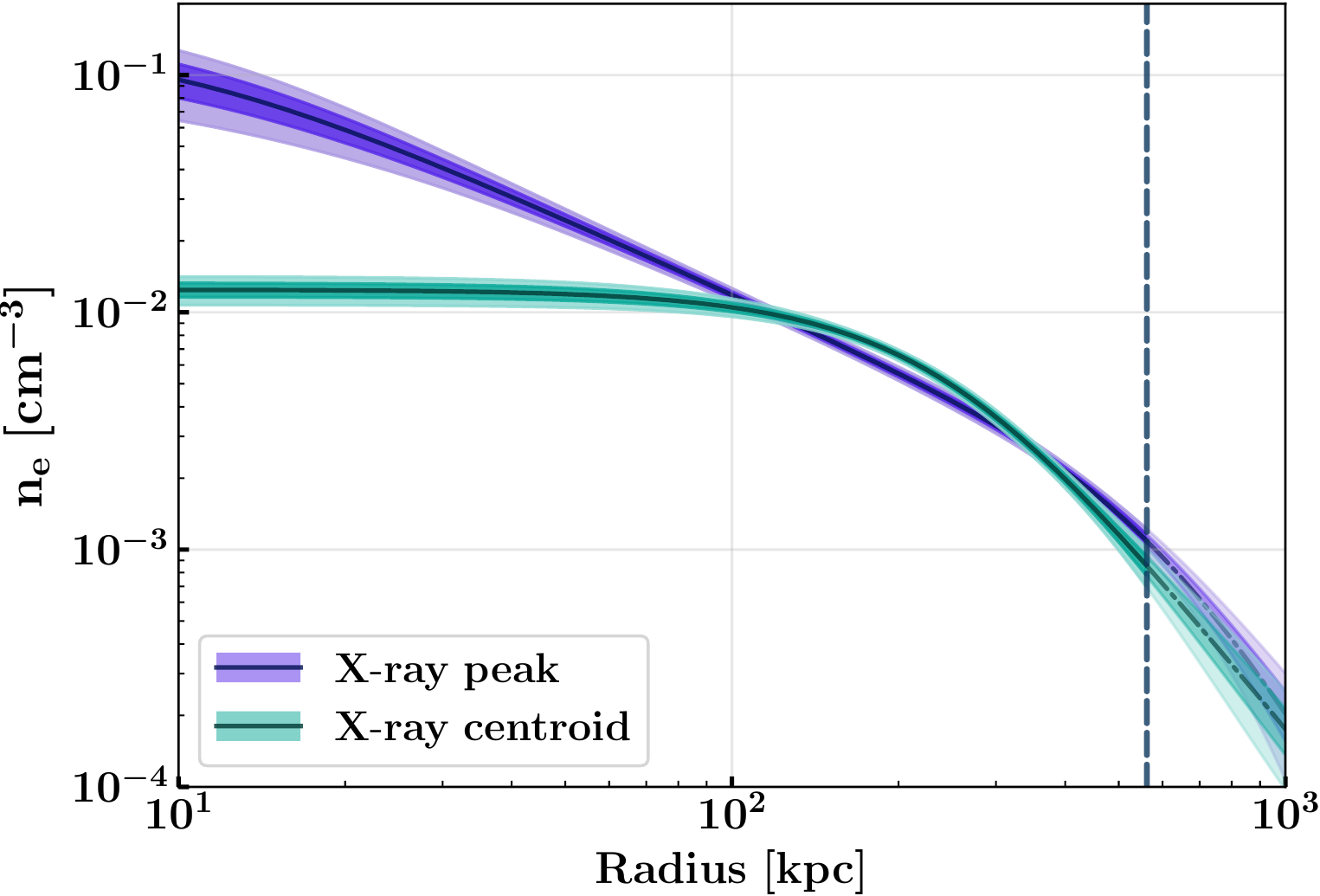}
\hspace{0.4cm}
\includegraphics[height=5.7cm]{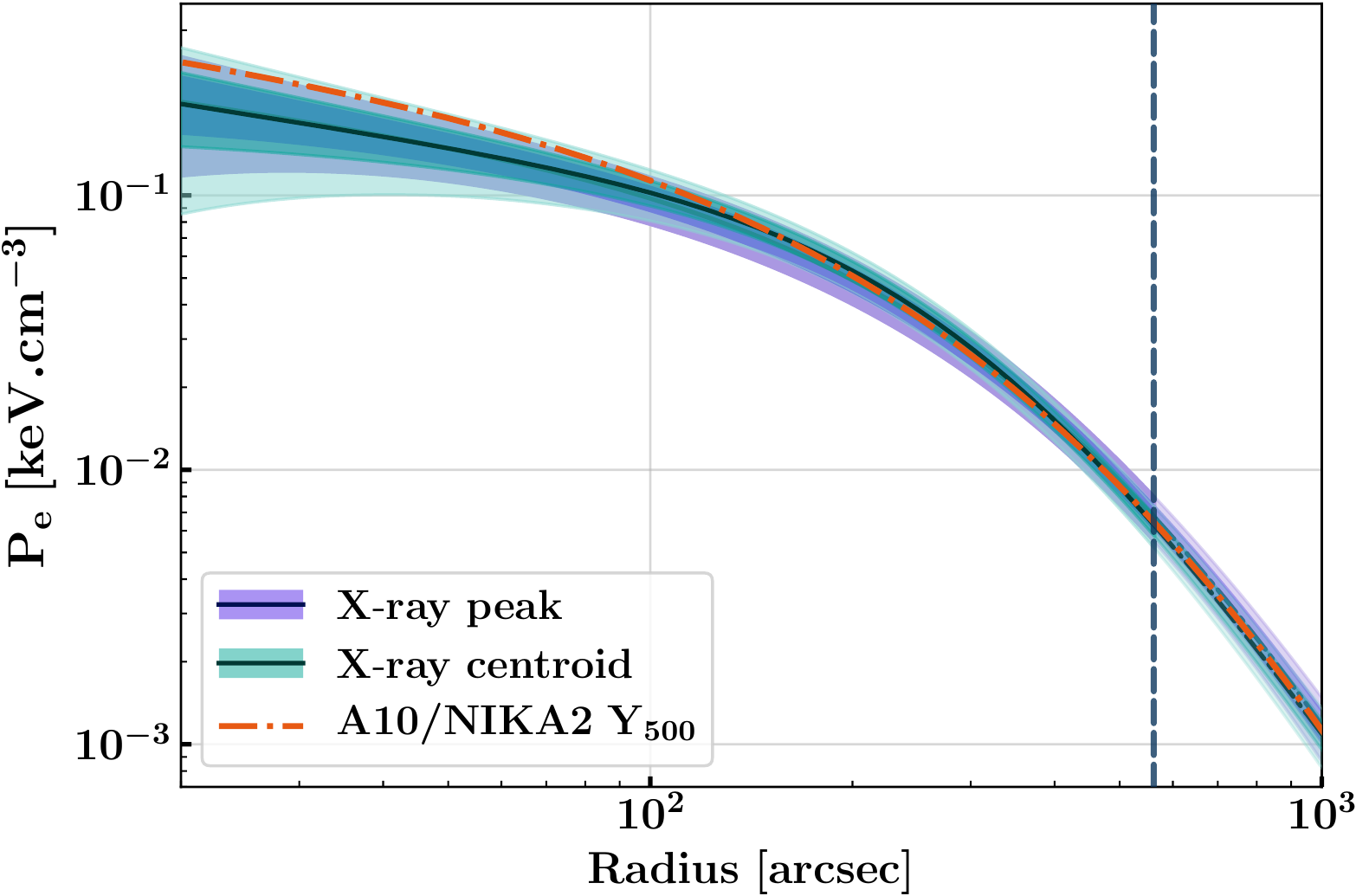}
\caption{{\footnotesize \textbf{Left:} Density profiles estimated from the analysis of the \chandra\ X-ray emission measure profile using a simplified Vikhlinin parametric model \citep[SVM;][]{vik06}. \textbf{Right:} Pressure profiles obtained from the deprojection of the NIKA2 SZ observations using a generalized Navarro-Frenk-White model \citep[gNFW;][]{nag07}. The orange profile gives the expected pressure profile using the universal pressure profile and scaling relation of \cite{arn10} along with the NIKA2 estimate of $Y_{500}$.  In both panels, the deprojection center is chosen to be the X-ray peak (purple) or the X-ray centroid (green). The dark lines give the NIKA2 and \chandra\ best-fits and the dark and light colored regions represent the 68\% and 95\% confidence regions on the ICM radial profiles. We highlight the position of the $3\sigma$ detection radius of $562~\mathrm{kpc}$ in the NIKA2 and \chandra\ data with a vertical dashed blue line and show the best-fit extrapolations in the outer regions with dash-dotted lines.}}
\label{fig:1d_P_ne}
\end{figure*}

This section is dedicated to the analysis of the radial distributions of the ICM thermodynamic properties of \moo. We use the complementarity of the \chandra\ X-ray data and the NIKA2 SZ observations in order to estimate both the cluster electron density profile and its pressure profile. This allows us to precisely measure the temperature and entropy distributions within the ICM under the assumption of spherical symmetry. We also study the effect of the cluster ellipticity and disturbed dynamics on the mass profile estimated under the assumption of hydrostatic equilibrium.

\subsection{\chandra\ density profile}\label{subsec:ne_prof}

We first describe the methodology followed to estimate the ICM density profile and present our results.

\subsubsection{Method}

We measure the ICM electron density profile $n_e(r)$ from the cluster emission integral given by:
\begin{equation}
\mathrm{EI} \equiv \int n_e n_p \, dV
\end{equation}
where  $n_p$ is the proton density within the ICM, and $V$ is the volume. The emission integral is related to the cluster surface brightness profile through the cooling function computed in the same energy band, $\Lambda (T,Z)$, depending on the ICM temperature $T$ and metallicity $Z$. We take into account the effects of the Galactic absorption and the variations of the effective area as a function of energy and position in the field of view to measure the cooling function. This is done by computing the normalization factor of the APEC model associated with the cluster spectrum in each annulus considered for the extraction of the surface brightness profile. This normalization factor is associated with a count rate $R$ measured in the annulus of area $A$ and it depends on both the ICM temperature and the effective area at this position. It is given by:
\begin{equation}
\mathrm{norm}(R,A) = \frac{10^{-14}}{4\pi [d_A (1+z)]^2} \int n_e n_p \, d\Omega dl
\end{equation} 
where $d_A$ is the angular diameter distance, and $d\Omega$ and $dl$ are the solid angle and line of sight differential elements respectively. We therefore compute the conversion coefficient between emission integral and surface brightness at each projected distance from the cluster center. We further correct this estimate for the spatial variations of the ICM temperature assuming a universal temperature profile from \cite{vik06} normalized to the mean spectroscopic temperature $T_X$ (see Sect. \ref{subsec:moo_xray}). This profile is represented as a black line in the right panel of Fig.~\ref{fig:chandra_1stprod}. We apply the conversion coefficient to the X-ray surface brightness profiles extracted on both the X-ray peak and the large-scale centroid in order to obtain the associated emission measure profiles:
\begin{equation}
\mathrm{EM}(r) = \int n_e n_p \, dl
\label{eq:em_prof}
\end{equation}

The ICM electron density profile $n_e(r)$ is estimated from the cluster emission measure profile using a forward fitting Bayesian framework. The cluster electron density distribution is modeled by a simplified Vikhlinin parametric model \citep[SVM;][]{vik06}, given by:
\begin{equation}
        n_e(r) = n_{e0} \left[1+\left(\frac{r}{r_c}\right)^2 \right]^{-3 \beta /2} \left[ 1+\left(\frac{r}{r_s}\right)^{\gamma} \right]^{-\epsilon/2 \gamma},
\label{eq:SVM}
\end{equation}
where $n_{e0}$ is the central density of the ICM,  $r_c$ is the core radius, and the inner and outer slopes of the profile are given by $\beta$ and $\epsilon$. The $\gamma$ parameter gives the width of the transition located at a radius $r_s$ at which an additional steepening in the profile occurs. The value of $\gamma$ is fixed at three since it provides a good fit of all the cluster profiles considered in the analysis of \cite{vik06b}. We assume the ionization fraction of a fully ionized plasma with an abundance of $0.3 Z_{\odot}$, \emph{i.e.}~$n_e/n_p = 1.199$ \citep{and89} in order to estimate the proton density profile in Eq. (\ref{eq:em_prof}). We perform a Markov chain Monte Carlo (MCMC) analysis to estimate the best-fit value of the five free parameters of the SVM model given the cluster emission measure profiles measured on both the X-ray peak and the large-scale centroid. This allows us to estimate the best-fit parameters that maximize the following Gaussian likelihood function:
\begin{equation}
\begin{tabular}{rl}
$-2 \mathrm{ln} \, \mathscr{L}_{\mathrm{X-ray}}$ & $ = \chi^2_{CXO} $\\[0.2cm]
&$= \sum_{i=1}^{N_{\mathrm{bin}}} [(EM_{CXO} - \widetilde{EM}) /  \Delta EM_{CXO}]^2_i$
\end{tabular}
\end{equation}
where $N_{\mathrm{bin}}$ is the number of bins in the emission measure profile $EM_{CXO}$ estimated from the \chandra\ data with uncertainties given by $\Delta EM_{CXO}$ and $\widetilde{EM}$ is the model of the emission measure profile obtained from the integration of the SVM profile along the line of sight. The best-fit electron density profile along with its error bars is estimated from the Markov chains after ensuring their convergence and applying a burn-in cutoff discarding half of the samples at the beginning of each chain.

\subsubsection{Results}\label{subsubsec:x_res}

The density profiles estimated from the analysis of the emission measure profiles obtained on the X-ray peak and the X-ray large-scale centroid are shown in the left panel of Fig.~\ref{fig:1d_P_ne} in purple and green, respectively. The density distributions are very well constrained from the cluster core up to $0.7R_{500}$. The density values at radii larger than ${\sim} 600$~kpc are obtained without significant constraints. We highlight this in the figure using a vertical dashed line showing the $3\sigma$ detection radius in the \chandra\ data. As the density estimates obtained beyond this radius are only extrapolations from the SVM model, we choose to focus our study in the radius range $r < 0.7R_{500}$. We observe a significant impact of the choice of deprojection center on the estimate of the ICM density distribution in the cluster core. There is an order of magnitude difference between the density estimates measured at 10~kpc from the X-ray peak and the X-ray large-scale centroid of \moo. This is due to the peculiar morphology of the cluster hosting its highest density core at ${\sim}100$~kpc from the large-scale centroid (see Sect. \ref{sec:moo_morphology}). Both density profiles are however fully compatible at $r \gtrsim 300$~kpc from the chosen deprojection center.\\

These results show the limit of a 1D analysis to describe the core dynamics of disturbed galaxy clusters. Indeed, if we assume the cuspiness $\alpha \equiv {|}d\,\mathrm{log}(n_e)/d\,\mathrm{log}(r){|}$ to be a good proxy to estimate the core dynamics of \moo\ \citep{vik07}, this cluster can either host a cool-core ($\alpha_{\mathrm{peak}} = 0.95$) or a disturbed core ($\alpha_{\mathrm{centroid}} = 0.04$) depending on the deprojection center that we choose. In the following, we will consider the profiles measured with respect to the X-ray peak to describe the core dynamics, and the ones centered on the large-scale centroid to estimate the integrated quantities such as $Y_{500}$.

\subsection{NIKA2 pressure profile}\label{subsec:pe_prof}

The \chandra\ data do not allow us to estimate the ICM pressure profile of \moo\ because the photon statistics are too low to measure a well defined temperature profile using X-ray spectroscopy (see Sect. \ref{subsec:moo_xray}). However, the NIKA2 SZ map at 150~GHz enables measuring directly the pressure distribution of the cluster (see Sect. \ref{subsec:tSZ_effect}). In this section, we describe the analysis procedure that we used to constrain the pressure profile and present our results. 

\subsubsection{Method}

We use the pipeline developed for the NIKA2 SZ large program \citep{per18} and detailed in \cite{rup18} in order to estimate the  pressure profile of \moo. The cluster pressure distribution is modeled by a generalized Navarro-Frenk-White model \citep[gNFW;][]{nag07}, given by:
\begin{equation}
        P_e(r) = \frac{P_0}{\left(\frac{r}{r_p}\right)^c \left(1+\left(\frac{r}{r_p}\right)^a\right)^{\frac{b-c}{a}}},
\label{eq:gNFW}
\end{equation}
where $a$ defines the width of the transition between the inner and the outer slopes $c$ and $b$, $P_0$ is a normalization constant, and $r_p$ is a characteristic radius. All these parameters are freely varying in an MCMC analysis performed to estimate the best-fit pressure profile of \moo\ given the NIKA2 SZ map at 150~GHz and the integrated Compton parameter measured by CARMA, \emph{i.e.}~$Y_{500} = (9.7 \pm 1.3) \times 10^{-5}~\mathrm{Mpc}^2$ \citep{gon15}.\\
\begin{figure*}[t]
\centering
\includegraphics[height=6.1cm]{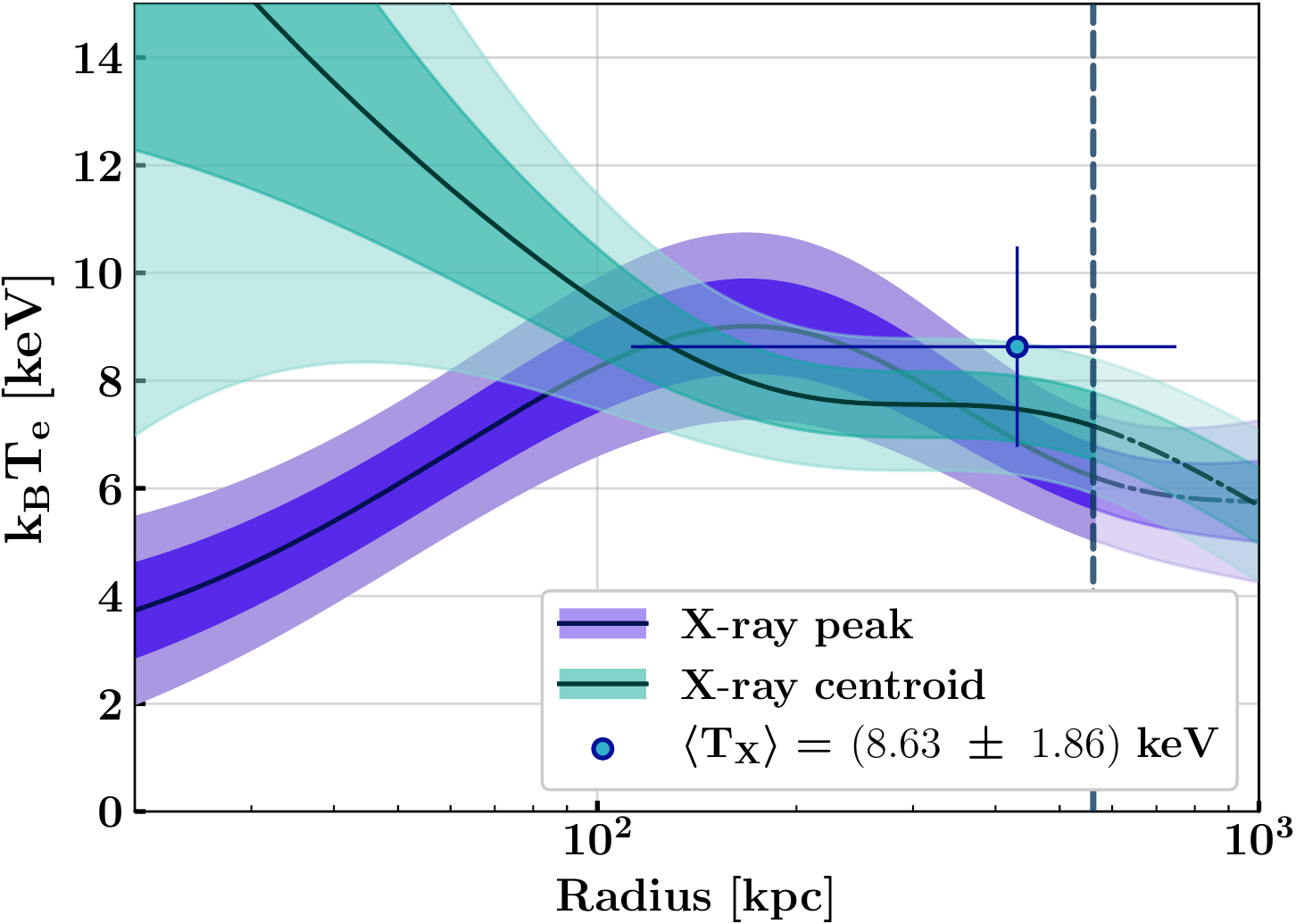}
\hspace{0.4cm}
\includegraphics[height=6.1cm]{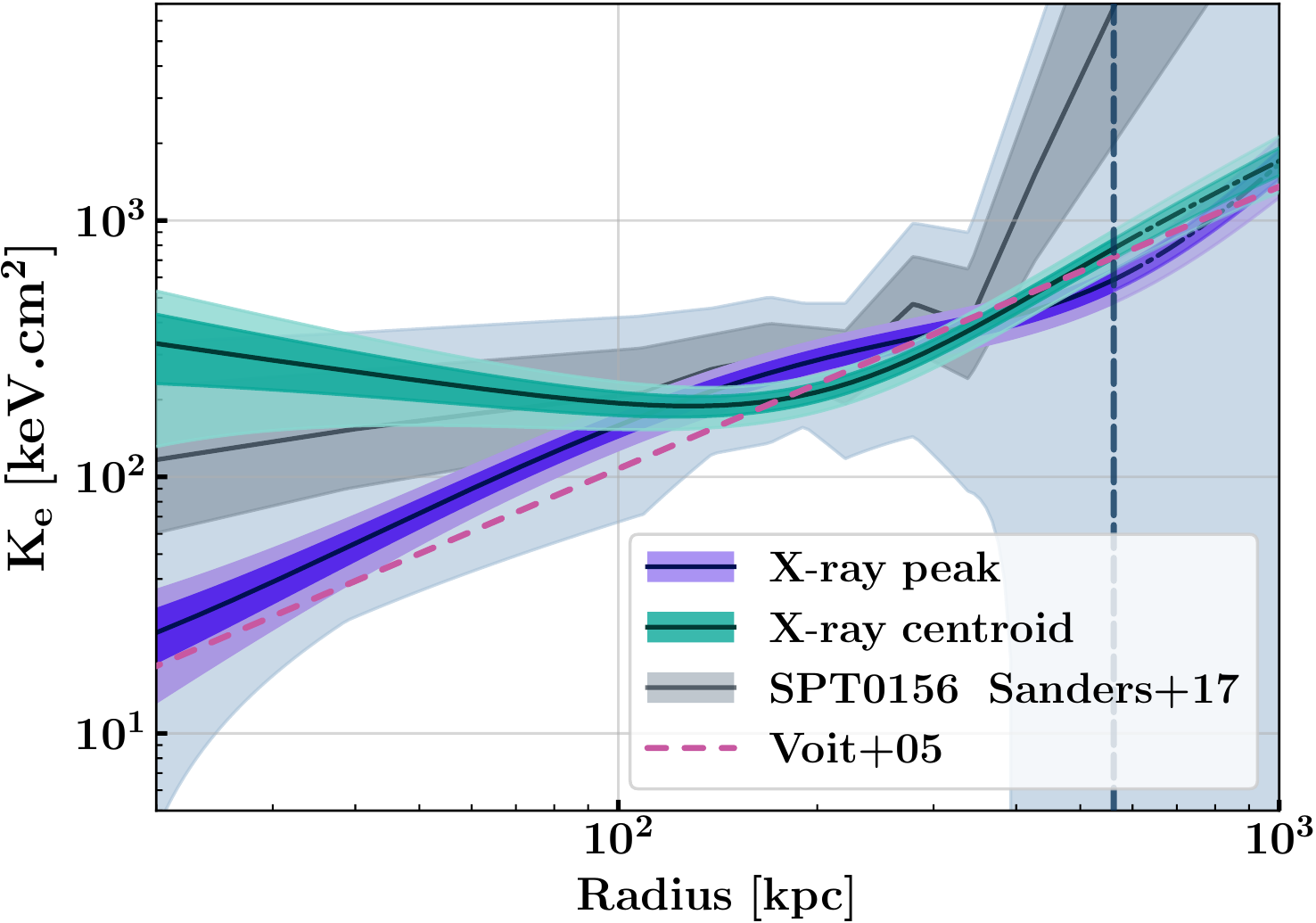}
\caption{{\footnotesize \textbf{Left:} Temperature profiles estimated from Eq. (\ref{eq:temp_entro}) using the combination of the NIKA2 pressure profiles and \chandra\ density profiles. The temperature radial distributions are compared with the single point temperature measurement from the \chandra\ spectroscopy data in the radius range $0.15 R_{500}<r<R_{500}$. \textbf{Right:} Entropy profiles estimated using the results of the NIKA2 and \chandra\ analyses. We compare our results with the entropy profile of the cluster $\mathrm{SPT\,CLJ0156}${-}$5541$ at redshift $z=1.22$ (in grey) estimated from a \chandra-only analysis based on a slightly higher number of X-ray photon counts \citep{san17}. We also display the self-similar expectation computed from the non-radiative simulation of \cite{voi05b} with a pink dashed line. In both panels, we use the same color code as in Fig.~\ref{fig:1d_P_ne} to distinguish the profiles estimated using the X-ray peak and the X-ray centroid as deprojection center. The unconstrained regions are shown as in Fig.~\ref{fig:1d_P_ne}.}}
\label{fig:1d_T_K}
\end{figure*}
At each step of the analysis, the pressure profile is integrated along the line of sight to obtain a Compton parameter map model. The latter is convolved with a 17.7~arcsec FWHM Gaussian kernel and with the analysis transfer function (see Sect. \ref{subsec:nika2_data}) to account for the small and large scale signal filtering in the NIKA2 data. The filtered Compton parameter map is converted into a SZ surface brightness map model using a conversion coefficient obtained by integrating the SZ spectrum within the NIKA2 bandpass at 150~GHz. Relativistic corrections are applied to the SZ spectrum based on the results of \cite{ito98} using a temperature profile obtained by combining the \chandra\ density profile (see Sect. \ref{subsec:ne_prof}) and the pressure profile. We model the radio source signal with a 2D Gaussian model centered at the position found in the FIRST survey. We fix the Gaussian FWHM to the NIKA2 angular resolution at 150~GHz and let the amplitude $\tilde{F}$ vary within a Gaussian prior based on the estimate obtained from the SED fit in Sect. \ref{sec:radio_pts}. The SZ map model $\tilde{M}$, obtained from the sum of the cluster and radio source signals, and the corresponding integrated Compton parameter $\tilde{Y}$ are finally compared with the NIKA2 and CARMA data using the following Gaussian likelihood:
\begin{equation}
\begin{tabular}{c@{\hskip -0.05mm}l}
        $-2 \mathrm{ln} \, \mathscr{L}_{\mathrm{SZ}}$&$ =\chi^2_{\mathrm{NIKA2}} + \chi^2_{\mathrm{CARMA}} + \chi^2_{\mathrm{radio}}$\\[0.2cm]
         &$ =\sum_{i=1}^{N_{\mathrm{pixels}}^{\mathrm{NIKA2}}} [(M_{\mathrm{NIKA2}} - \tilde{M})^T C_{\mathrm{NIKA2}}^{-1} (M_{\mathrm{NIKA2}} - \tilde{M})]_i $ \\[0.2cm]
         &$ + \left[\frac{Y_{\mathrm{500}} - \tilde{Y}}{\Delta Y_{\mathrm{500}}}\right]^2 + \left[\frac{F_{\mathrm{150~GHz}} - \tilde{F}}{\Delta F_{\mathrm{150~GHz}}}\right]^2$
\label{eq:chi2_NK2_CARMA}
\end{tabular}
\end{equation}
where $M_{\mathrm{NIKA2}}$ and $C_{\mathrm{NIKA2}}$ are the NIKA2 SZ surface brightness map and noise covariance matrix at 150~GHz, and $Y_{\mathrm{500}}$ is the CARMA integrated Compton parameter measured with the uncertainty $\Delta Y_{\mathrm{500}}$. As the raw data analysis method detailed in Sect. \ref{subsec:nika2_data} has allowed us to measure an almost flat noise power spectrum in the NIKA2 map at 150~GHz, the pixel-to-pixel correlation induced by residual correlated noise in the NIKA2 map is negligible compared to the RMS noise within each pixel at this frequency. We therefore assume that the NIKA2 $\chi^2$ is given by:
\begin{equation}
\chi^2_{\mathrm{NIKA2}} = \sum_{i=1}^{N_{\mathrm{pixels}}^{\mathrm{NIKA2}}} [(M_{\mathrm{NIKA2}} - \tilde{M}) / M_{\mathrm{RMS}}]^2_i
\end{equation}
where $M_{\mathrm{RMS}}$ is the map of the RMS noise in each pixel of the NIKA2 map at 150~GHz. It is computed from simulated noise maps including the instrumental, atmospheric and astrophysical contaminants (see Sect. \ref{subsec:moo_sz}). The fact that we do not need to use a noise covariance matrix at 150~GHz to account for the residual noise properties at this frequency allows us to decrease the computation time of the likelihood value at each step of the MCMC from 0.2~s to 0.5~ms. We compute the best-fit pressure profile and its associated error bars from the samples remaining in the Markov chains after their convergence and a burn-in cutoff discarding the first 50\% of each chain.

\subsubsection{Results}

The pressure profiles obtained by using both the X-ray peak and the X-ray large-scale centroid as deprojection centers are shown in the right panel of Fig.~\ref{fig:1d_P_ne} in purple and green, respectively. As the SZ signal has been mapped by NIKA2 up to similar angular scales as the X-ray signal measured by \chandra, we are also able to estimate the pressure profile from the cluster core up to $0.7R_{500}$. As explained in Sect. \ref{subsubsec:x_res}, we distinguish the results obtained at larger radii using dash-dotted lines because they are overconstrained by the model. The two pressure profiles estimated from this analysis are fully compatible although the deprojection centers are quite different. This comes from the fact that the highest SZ signal amplitude is measured over a very extended region that encloses both the X-ray peak and the X-ray large-scale centroid (see Sect. \ref{sec:moo_morphology}). For this reason, the inner slope of the pressure profile does not depend significantly on the deprojection center if the latter is chosen within the closed region delimited by the orange line in Fig.~\ref{fig:rgb_map}.\\

In contrast, the flux of the radio source has a significant impact on the pressure profile estimate in the cluster core. If we consider the X-ray large-scale centroid as a deprojection center, the SZ signal is almost azimuthally symmetric and a gNFW model provides a good description of the cluster morphology observed in the NIKA2 map at 150~GHz. The best-fit value of the radio source flux at the end of the MCMC is found to be $\tilde{F} = (1.6 \pm 0.1)$~mJy, which is consistent with the expected flux from the source SED (see Sect. \ref{sec:radio_pts}). However, if we use the X-ray peak as a deprojection center, the SZ signal is not azimuthally symmetric because of the overpressure region located to the west of the X-ray peak. Furthermore, in this case, the SZ peak position in the map model matches the one of the radio source. For these reasons, the radio source flux tends to be overestimated in order to maximize the SZ signal amplitude to its west in the SZ map model. Thus, the best-fit value of the radio source flux in this analysis is found to be $\tilde{F} = (2.2 \pm 0.2)$~mJy which is more than $2\sigma$ higher than the expected flux from the source SED. In order to avoid biasing the estimate of the pressure distribution in the cluster core when using the X-ray peak as a deprojection center, we fix the radio source flux to its expected value, \emph{i.e.}~$\tilde{F} = 1.5$~mJy in the corresponding MCMC. The inner slope of the estimated profile (in purple in Fig.~\ref{fig:1d_P_ne}) is then fully compatible with the one measured using the X-ray large-scale centroid.\\

We compare our pressure profile estimates with the universal pressure profile from \cite{arn10} scaled to the same integrated Compton parameter (see Sect. \ref{subsec:mass_prof}). The profile is shown with the orange dash-dotted line in the right panel of Fig.~\ref{fig:1d_P_ne}. We do not find any significant deviations from the shape of the universal pressure profile except for a slightly shallower inner slope. Although a single cluster analysis cannot be used to draw conclusions on the redshift evolution of the ICM pressure distribution, it is interesting to note that the pressure profile of this disturbed cluster at very high redshift is compatible with the mean profile estimated from the study of an X-ray selected sample at $z < 0.2$ \citep{arn10}.

\subsection{Gas temperature, entropy, and cooling time}\label{subsec:T_K_prof}

\begin{figure*}[t]
\centering
\includegraphics[height=5.7cm]{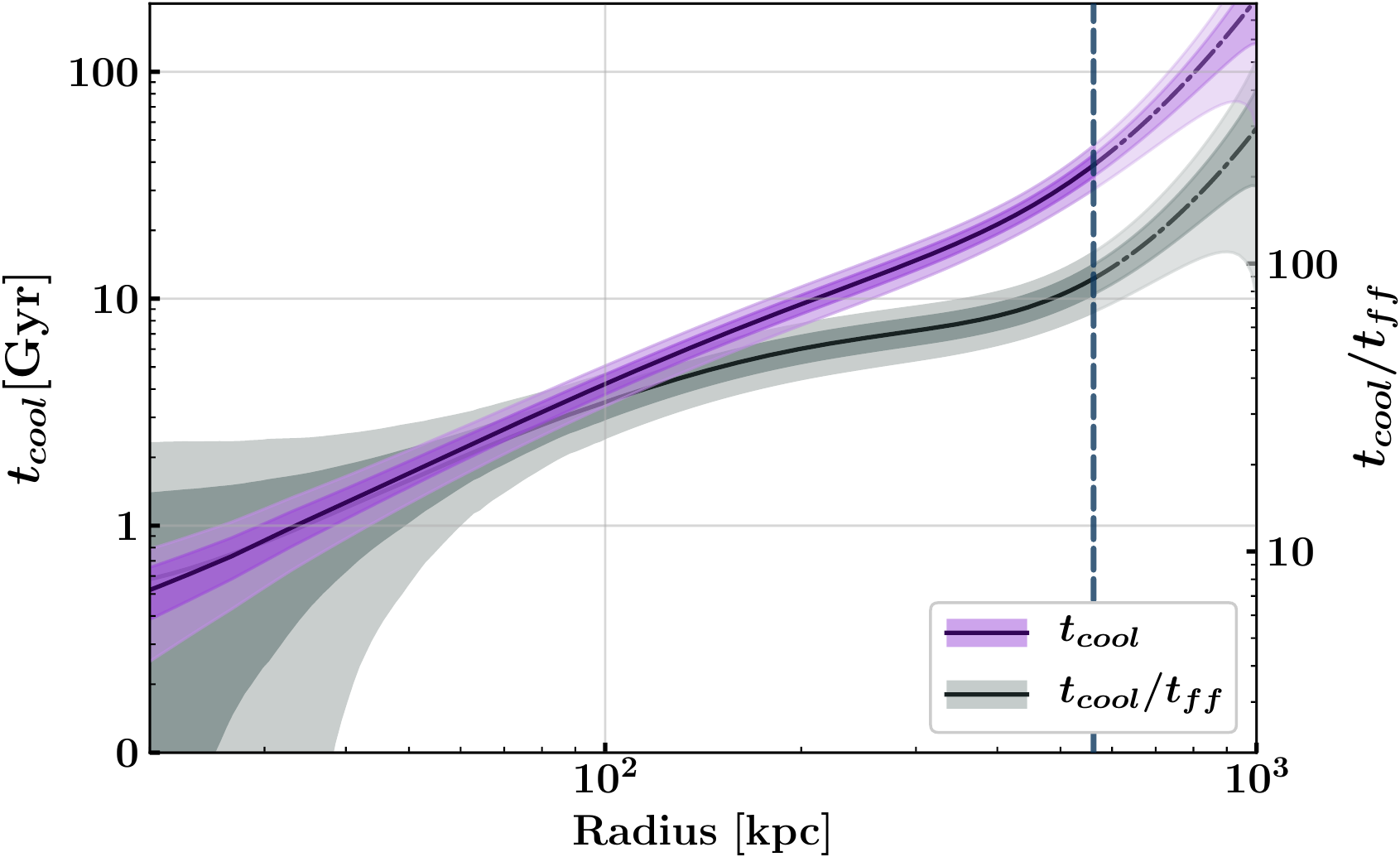}
\hspace{0.4cm}
\includegraphics[height=5.7cm]{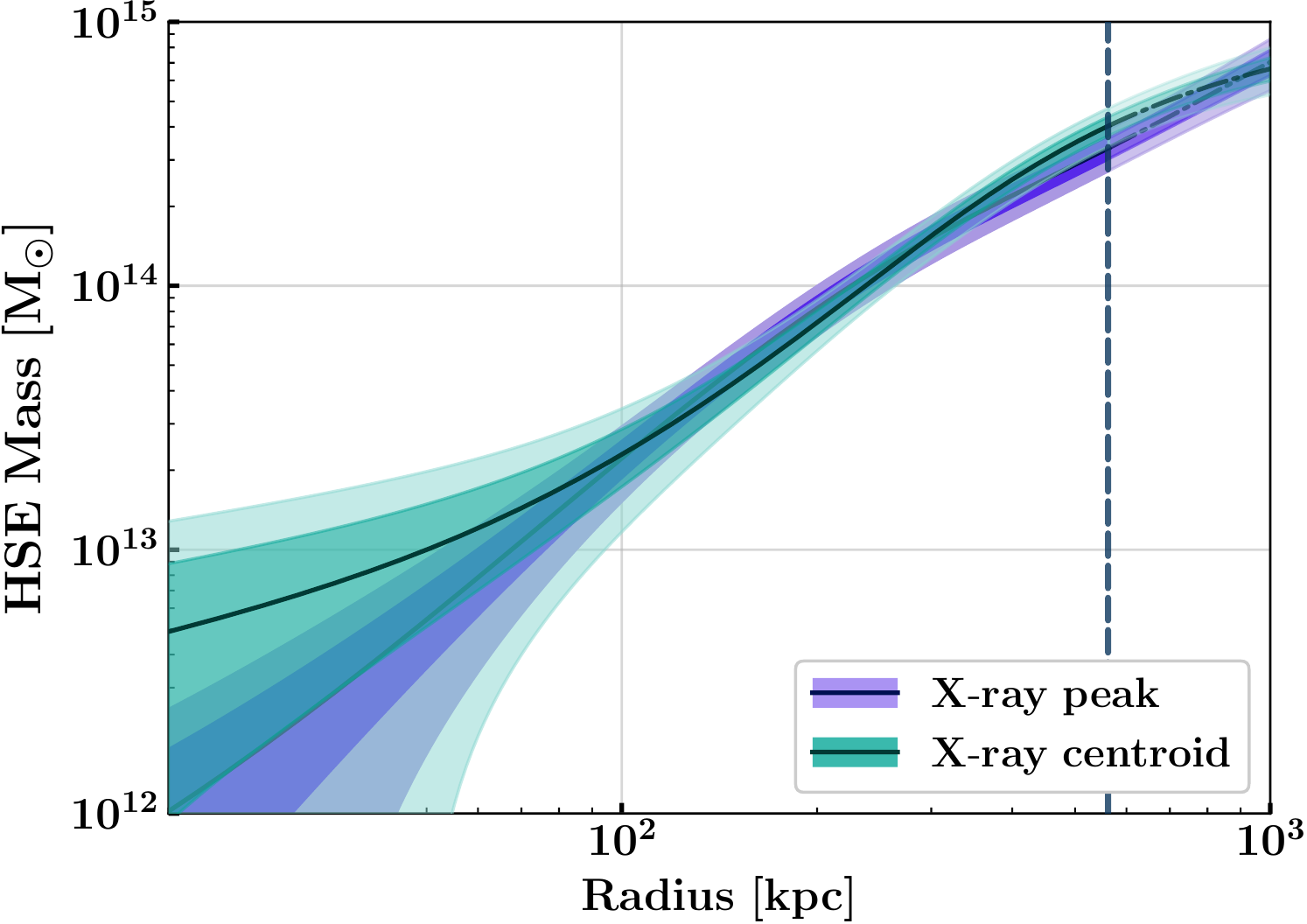}
\caption{{\footnotesize \textbf{Left:} Profiles of the ICM cooling time (magenta) and of the ratio of the cooling to free-fall times (grey) estimated from the density, temperature, and hydrostatic mass profiles issued from the joint analysis of the NIKA2 and \chandra\ data using the X-ray peak as deprojection center. \textbf{Right:} Hydrostatic mass profiles estimated from Eq. \ref{eq:hse_mass} using the density and pressure profiles deprojected with respect to the X-ray peak (purple) and the X-ray centroid (green). In both panels, the black line gives the best-fit estimate and the dark and light regions show the $1\sigma$ and $2\sigma$ confidence regions. The unconstrained regions are shown as in Fig.~\ref{fig:1d_P_ne}.}}
\label{fig:1d_M_tcool}
\end{figure*}

The combination of the \chandra\ density profile and NIKA2 pressure profile allows us to estimate the other ICM thermodynamic properties of \moo\ without relying on X-ray spectroscopy.\\

Under the ideal gas assumption, the ICM temperature and entropy profiles are estimated from the following equations:
\begin{equation}
k_B T_e(r) = \frac{P_e(r)}{n_e(r)}~~~\mathrm{and}~~~K_e(r) =  \frac{P_e(r)}{n_e(r)^{5/3}},
\label{eq:temp_entro}
\end{equation}
where, $k_B$ is the Boltzmann constant. We show the temperature and entropy profiles obtained from the combination of the \chandra\ density profile (see Sect. \ref{subsec:ne_prof}) and NIKA2 pressure profile (see Sect. \ref{subsec:pe_prof}) in Fig.~\ref{fig:1d_T_K}. We use the same color scheme as the one considered in Fig.~\ref{fig:1d_P_ne} in order to differentiate the profiles measured using the X-ray peak (purple) and the X-ray large scale centroid (green) as deprojection centers.\\

The temperature profiles (left panel) estimated with these two deprojection centers are significantly different at radii $r \lesssim 100$~kpc. We measure a monotonically decreasing temperature from the X-ray centroid up to $0.7R_{500}$. The central temperature around the SZ peak reaches high values around 14~keV within the enclosed region defined in Fig.~\ref{fig:rgb_map} (orange  line). However, the profile estimated by considering the X-ray peak has the typical shape expected for a cool-core cluster. The temperature increases from a low central value of ${\sim}4$~keV up to a maximum of $9$~keV at $170$~kpc from the X-ray peak. It then decreases at larger radii. These profiles estimated from the combination of X-ray and SZ results are also compatible with the mean ICM temperature estimated from the \chandra\ spectroscopic data (blue point). We emphasize the large leap forward that such a joint analysis of SZ and X-ray data allows us to make on the characterization of the ICM temperature profile at $z>1$. Indeed, having a precise measurement of the central temperature, the inner slope of the profile, and the location of the maximum temperature value with X-ray spectroscopy would require an increase of the \chandra\ exposure of at least an order of magnitude.\\
\begin{table*}[t]
\begin{center}\caption{{\footnotesize Estimates of the $R_{500}$ radius, hydrostatic mass ($M_{500}$), and integrated Compton parameter ($Y_{500}$) computed from the combination of the \chandra\ density and NIKA2 pressure profiles. The values estimated by considering both the X-ray centroid and peak emission as deprojection centers are compared.}}\label{tab:int_param}\begin{tabular}{ccc}
\hline
\hline
 & \textbf{X-ray centroid} & \textbf{X-ray peak} \\
\hline
$R_{500}$ [kpc] & $841 \pm 31$ & $812 \pm 41$\\
$M_{500}~[\mathrm{\times 10^{14}~M_{\odot}}]$ & $6.06 \pm 0.68$ & $5.49 \pm 0.85$ \\
$Y_{500}~[\mathrm{\times 10^{-4}~arcmin^2}]$ & $2.95 \pm 0.21$ & $2.96 \pm 0.23$ \\
\hline
\hline
\end{tabular}
\end{center}
\end{table*}

The entropy profiles estimated from the combination of the NIKA2 and \chandra\ results are shown in the right panel of Fig.~\ref{fig:1d_T_K}. The outer slopes of both profiles are compatible with the self-similar expectation obtained from non-radiative simulations and shown with a pink dashed line \citep{voi05b}. However, the radial distribution of the gas entropy  estimated by using the X-ray peak is significantly different from the one obtained with the large-scale centroid at $r \lesssim 100$~kpc. We measure a flat entropy distribution at a high value of ${\sim}230~\mathrm{keV\cdot cm^2}$ around the X-ray centroid. This is consistent with the disturbed ICM activity at the location of the SZ peak (see Sect. \ref{sec:moo_morphology}). On the other hand, the entropy profile estimated using the X-ray peak is monotonically increasing from a low central value of ${\sim}25~\mathrm{keV\cdot cm^2}$ at $20$~kpc from the deprojection center up to ${\sim}600~\mathrm{keV\cdot cm^2}$ at $0.7R_{500}$. This profile is consistent with the one expected for a cool-core cluster in which the central radiative cooling is balanced by the activity of the radio loud AGN detected within the BCG. The central entropy value estimated from this profile depends on the flux of the radio source considered in the NIKA2 MCMC analysis. A flattening of the radio source SED at high frequency \citep[\emph{e.g.}][]{hog15} would indeed result in a higher expected flux at 150~GHz and therefore to a higher central value of the deprojected pressure distribution from the NIKA2 data. However, we have checked that the central entropy value is always lower than ${\sim}70~\mathrm{keV\cdot cm^2}$ if we consider a point source flux lower than 4~mJy at 150~GHz. This shows that the existence of a cool-core at the position of the X-ray peak holds even if the radio source flux at frequencies $\nu > 100$~GHz remains at the same value measured by CARMA at 31~GHz.\\

Finally, we compare our constraints on the entropy profile of \moo\ with the results obtained by \cite{san17} from the analysis of \chandra\ data measured on the cluster $\mathrm{SPT\,CLJ0156}${-}$5541$ at $z=1.22$ (grey profile). A refined analysis based on the assumption of an underlying dark-matter potential enabled theses authors to estimate an entropy profile with a total of 2300 X-ray counts. However, the comparison of this profile with baseline entropy distributions extracted from simulations is difficult because of the large error bars associated with this estimate. Here, we show that a joint analysis combining a slightly lower number of X-ray counts (see Sect. \ref{sec:chandra_obs}) with high angular resolution SZ observations results in a highly constrained entropy profile even at $z>1$.\\

We further describe the properties of the cluster cool-core by computing the isochoric cooling time and free-fall time profiles given by:
\begin{equation}
t_{cool}(r) = \frac{3}{2} \frac{(n_e+n_p) \, k_B T_e}{n_e n_p \, \Lambda(T_e,Z)}~~~\mathrm{and}~~~t_{ff}(r) = \sqrt{\frac{2r}{g(r)}}
\end{equation}
where $g(r) = G \cdot B \cdot M_{HSE}(r) / r^2$ is the gravitational acceleration caused by the total mass $B \cdot M_{HSE}(r)$ within a sphere of radius $r$, and $G$ is the gravitational constant. The hydrostatic bias $B = 1/(1-b)$ is a correction applied to the hydrostatic mass profile $M_{HSE}(r)$ to account for the departure of the gas dynamical state from hydrostatic equilibrium. We use the same ionization fraction considered in Sect. \ref{subsec:ne_prof} in order to estimate the proton density profile $n_p(r)$ from the electron density distribution $n_e(r)$ obtained by considering the X-ray peak as a deprojection center. Furthermore, we use the cooling function estimated by \cite{sut93} for an optically-thin plasma with a $0.3Z_{\odot}$ metallicity to compute $\Lambda(T_e,Z)$ from the temperature profile centered on the X-ray peak, shown in Fig.~\ref{fig:1d_T_K}. The free-fall time is estimated by using the hydrostatic mass profile computed from the combination of the \chandra\ density and NIKA2 pressure profiles (see Sect. \ref{subsec:hse_ytot} below). We use an hydrostatic bias parameter fixed to $b = 0.2$, following the results of numerical simulations and the comparisons between weak lensing and SZ/X-ray cluster mass estimates \citep[see \emph{e.g.} Fig.~10 in][]{sal18}. The cooling time profile as well as the ratio $t_{cool}/t_{ff}$ are represented in the left panel of Fig.~\ref{fig:1d_M_tcool} in magenta and grey respectively. We measure a cooling time at 20~kpc from the X-ray peak location of $t_{cool} = (0.51\pm 0.13)$~Gyr. This is compatible with the distribution of central cooling times estimated by \cite{mcd13} on a sample of SPT-selected cool-core clusters. We note that this central cooling time is fairly low given the cluster redshift. This supports the results shown in Fig.~6 of \cite{mcd13} on the absence of significant redshift evolution of the gas cooling properties in cluster cores. We measure a ratio $t_{cool}/t_{ff} \simeq 9$ at the core of \moo. The relationship between the minimum value of $t_{cool}/t_{ff}$ and the central entropy is therefore consistent with the distribution measured on the sample of X-ray selected clusters at a mean redshift $\langle z \rangle = 0.17$ dubbed the Archive of Chandra Cluster Entropy Profile Tables \citep[ACCEPT;][]{voi15}. In summary, although \moo\ displays many features of a morphologically disturbed galaxy cluster (see Sect. \ref{sec:moo_morphology}) at a redshift $z = 1.2$, its core properties are very similar to those observed in typical cool-core clusters at low redshift.

\subsection{Hydrostatic mass and Compton parameter}\label{subsec:hse_ytot}

\begin{figure*}[t]
\centering
\includegraphics[height=7.9cm]{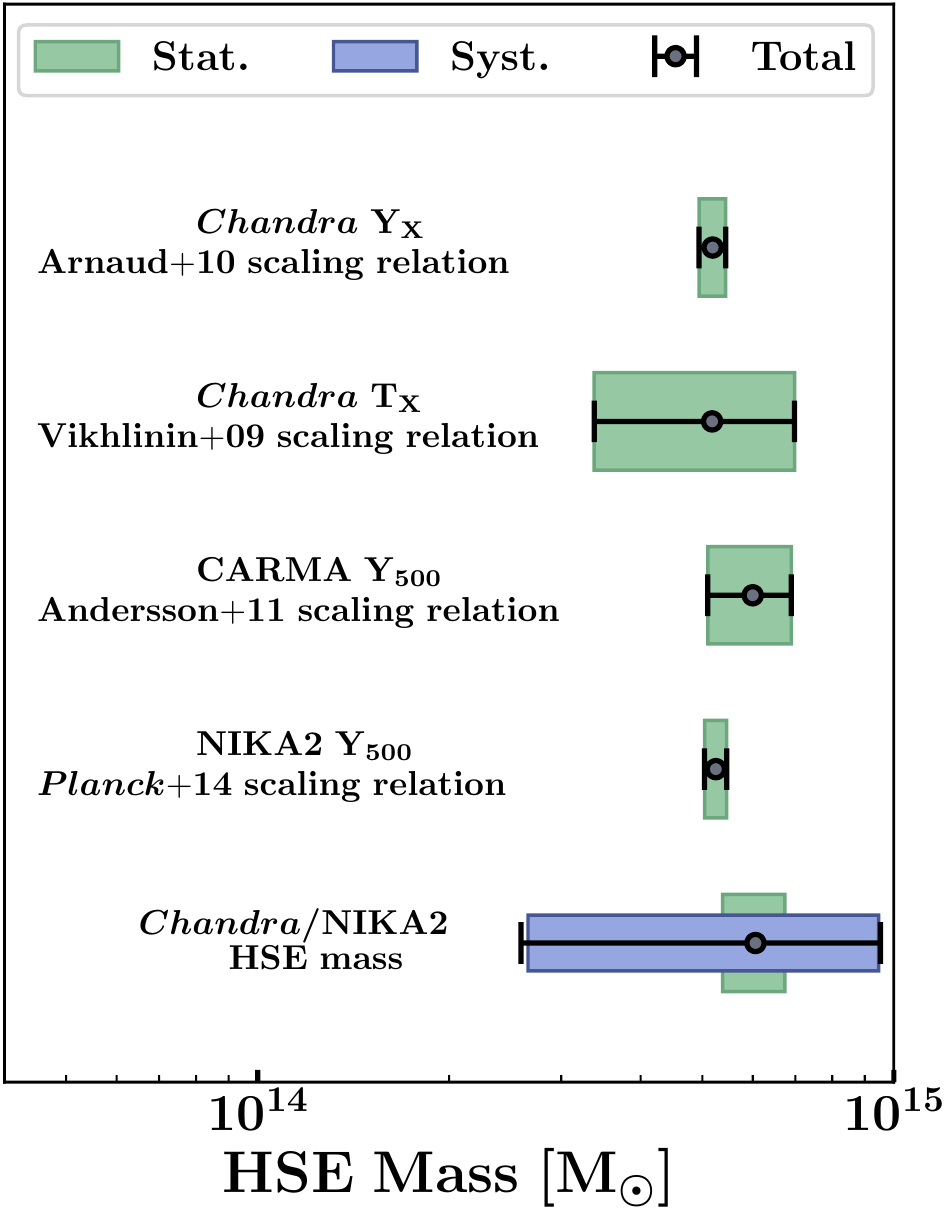}
\hspace{0.2cm}
\includegraphics[height=7.9cm]{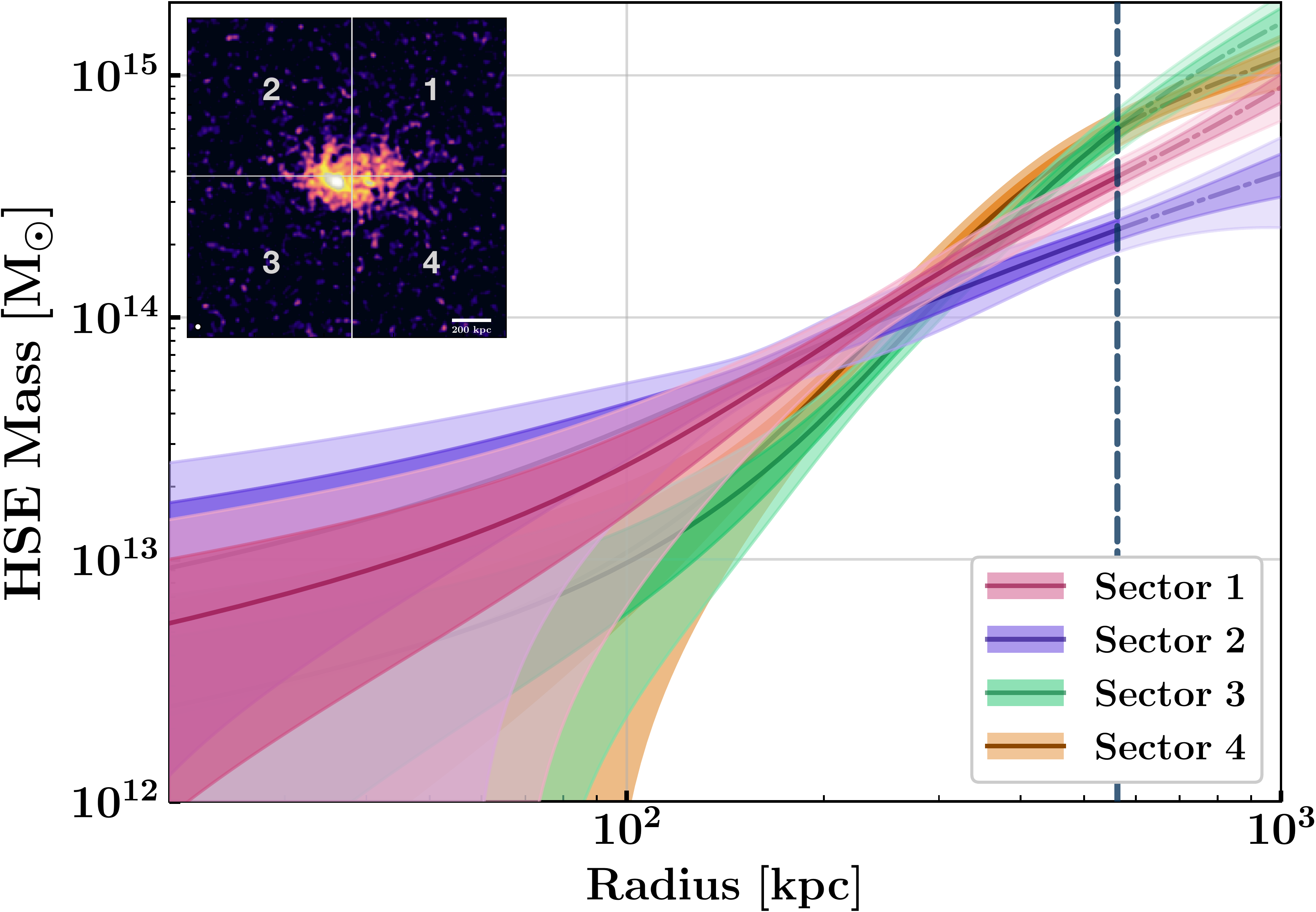}
\caption{{\footnotesize \textbf{Left:} Comparison of the mass estimates of \moo\ obtained from the \chandra\ $Y_X$ and $T_X$ observables and the CARMA/NIKA2 integrated Compton parameters $Y_{500}$, along with scaling relations calibrated at low redshift with the hydrostatic mass value estimated from the combination of the \chandra\ density profile and NIKA2 pressure profile centered on the X-ray centroid. Statistical uncertainties are shown in green while the systematic uncertainty on the hydrostatic mass estimate (in blue) is obtained from the scatter between the mass profiles shown in the right panel. \textbf{Right:} Hydrostatic mass profiles obtained from the combination of the NIKA2 pressure profiles and \chandra\ density profiles extracted in four different angular sectors centered on the large-scale X-ray centroid, shown in the inset map of the \chandra\ X-ray surface brightness. Each mass profile is labeled with a number corresponding to the considered angular sector. The unconstrained regions are shown as in Fig.~\ref{fig:1d_P_ne}.}}
\label{fig:hse_mass}
\end{figure*}

The estimates of both the density and pressure profiles of \moo\ allow us to infer the total mass of this cluster independently of any scaling relation calibrated at lower redshift. We estimate the total mass enclosed within a radius $r$ under the hydrostatic equilibrium assumption:
\begin{equation}
M_{\rm HSE}(r) = -\frac{r^2}{\mu_{gas} m_p n_e(r) G} \frac{dP_e(r)}{dr},
\label{eq:hse_mass}
\end{equation}
where $m_p$ is the proton mass and $\mu_{\rm{gas}} = 0.61$ is the mean molecular weight of the gas. The hydrostatic mass profiles  computed from the density and pressure profiles estimated by using the X-ray peak and the X-ray large-scale centroid as deprojection centers are both represented in the right panel of Fig.~\ref{fig:1d_M_tcool} in purple and green respectively. The profiles are compatible with each other from the cluster core up to $0.7R_{500}$ although the inner slope of the profile measured from the X-ray centroid is shallower than the one obtained at the X-ray peak. The hydrostatic mass profile of \moo\ is used to compute its density contrast profile $\langle \, \rho(r) \, \rangle/\rho_c$, where $\rho(r) = M_{\rm HSE}(r) / V(r)$, $V(r)$ is the cluster volume at a radius $r$, and $\rho_c$ is the critical density of the Universe at $z=1.2$. This allows us to measure directly the value of $R_{500}$. The estimates of $R_{500}$ measured from the two mass profiles enable us to compute the total mass of the cluster $M_{500}$ as well as its integrated Compton parameter $Y_{500}$. The results are  summarized in Tab. \ref{tab:int_param}.\\

The results obtained with the two mass profiles are all compatible within their $1\sigma$ confidence intervals. The error bars associated with the estimates derived from the mass profile measured using the X-ray peak are however larger than the ones obtained with the X-ray centroid. A spherical model is indeed ill-suited to subtract the extended emission to the west of the X-ray peak without creating strong residuals to its east in both \chandra\ and NIKA2 data. While using the X-ray centroid as deprojection center induces residuals around the X-ray peak because the X-ray signal is not azimuthally symmetric in this area, the extended emission is subtracted with more accuracy in this case. More quantitatively, the minimum $\chi^2$ value obtained at the end of the NIKA2 MCMC analysis (see Sect. \ref{subsec:pe_prof}) is decreased by 11\% if we use the X-ray centroid instead of the X-ray peak as deprojection center. Although considering the X-ray peak as deprojection center is mandatory to have an accurate description of the cluster core properties (see Sect. \ref{subsec:T_K_prof}), using the X-ray centroid leads to both more accurate and more precise estimates of the integrated quantities of \moo\ such as $M_{500}$ and $Y_{500}$. We therefore discuss the results obtained with the X-ray centroid in the following.\\

The $R_{500}$ radius measured with this multi-wavelength analysis is compatible with the estimate based on the measurement of the mean ICM spectroscopic temperature and the $M_{500}{-}T_X$ relation from \cite{vik09} (see Sect. \ref{subsec:chandra_data}). Moreover, our estimate of $M_{500}$ is fully compatible with the value determined by \cite{gon15} using the CARMA integrated Compton parameter along with the \cite{and11} $Y_{500}{-}M_{500}$ scaling relation, \emph{i.e.}~$M_{500}^{CARMA} = (6.0 \pm 0.9) \times 10^{14}~\mathrm{M}_{\odot}$. Based on our measurement of the integrated Compton parameter of \moo\ computed from the spherical integral of the NIKA2 pressure profile up to $R_{500}$, we estimate the expected mass of this cluster using the \planck\ scaling relation \citep{pla14}: $M_{500}^{Planck} = (5.25 \pm 0.21) \times 10^{14}~\mathrm{M}_{\odot}$. The cluster mass reported in Tab. \ref{tab:int_param} is therefore compatible with the scaling-relation-based value given the error bars of our estimate and the intrinsic scatter associated with the \planck\ scaling relation calibrated at redshifts $z<0.45$. In addition, we compute the pseudo-integrated Compton parameter $Y_X = M_{g,500} \times T_X$ using the gas mass measurement $M_{g,500}$ obtained by integrating the \chandra\ density profile centered on the X-ray centroid, and the mean spectroscopic temperature found in Sect. \ref{subsec:moo_xray}. We estimate the total mass of the cluster using the $M_{500}{-}Y_X$ scaling relation of \citep{arn10} and find $M_{500}^{Y_X} = (5.19 \pm 0.25) \times 10^{14}~\mathrm{M}_{\odot}$. We therefore confirm that \moo\ is the most massive galaxy cluster known to date at $z > 1.15$ from an analysis based on spherical models of the ICM density and pressure distributions and the assumption of hydrostatic equilibrium. We compare all mass estimates of this cluster given in this paper in the left panel of Fig.~\ref{fig:hse_mass}. We highlight that our hydrostatic mass estimate obtained by combining the deprojected ICM pressure and density profiles is compatible with the ones computed based on mass-observable scaling relation calibrated using low-redshift cluster samples.

\subsection{Systematic error on the hydrostatic mass}\label{subsec:mass_prof}

As shown in Sect. \ref{sec:moo_morphology}, \moo\ is not a relaxed cluster nor it can be considered as spherical given the precision of the \chandra\ and NIKA2 measurements. Thus, we perform a dedicated analysis in order to estimate a systematic uncertainty caused by both the expected departure of the cluster dynamical state from hydrostatic equilibrium and the modeling error induced by the use of a spherical model to deproject the ICM pressure and density distributions.\\

We analyze both the \chandra\ and NIKA2 observations in four angular sectors of $90^{\circ}$ opening angles centered on the large-scale X-ray centroid (see inset in the right panel of Fig.~\ref{fig:hse_mass}). The density and pressure profiles are estimated in each of these sectors following the methodology described in Sect. \ref{subsec:ne_prof} and \ref{subsec:pe_prof}. We use Eq. (\ref{eq:hse_mass}) to compute the mass profiles based on these results assuming that the ICM is in hydrostatic equilibrium in each sector and that its thermodynamic properties can be described by spherically symmetric distributions. If the cluster was perfectly spherical and morphologically relaxed, these four mass profiles would be compatible at all scales. We represent the hydrostatic mass profiles estimated in each sector in Fig.~\ref{fig:hse_mass}. Although these profiles are compatible in the cluster center, we measure significant discrepancies at radii larger than ${\sim 400}$~kpc. We consider these differences as an indicator of the systematic error associated with the estimate of the total mass of the cluster. As the outer slopes and amplitudes of the four profiles are different, the values of $R_{500}$ estimated from the radial distribution of the density contrast in each sector are also significantly different. Therefore, we choose to define the systematic uncertainty associated with our estimate of the hydrostatic mass of this cluster as the standard deviation between the mass values measured with each profile at a radius of 841~kpc, \emph{i.e.}~the $R_{500}$ estimate given in Tab. \ref{tab:int_param}. We also measure the mean of these four mass values to define a new estimate of the cluster total mass. The hydrostatic mass of \moo\ obtained by this analysis is given by $M_{841~\mathrm{kpc}} = (7.4 \pm 3.4) \times 10^{14}~\mathrm{M}_{\odot}$. Although this result is fully compatible with the hydrostatic mass estimate given in Tab. \ref{tab:int_param}, we emphasize that the dispersion of the mass estimates between the four angular sectors is three times higher than the error bar associated with the mass measurement at $841~\mathrm{kpc}$ in each sector. Our final estimate of the hydrostatic mass of \moo\ is given by the measurement referenced in Tab. \ref{tab:int_param} along with this additional systematic uncertainty: $M_{500} = (6.06 \pm 0.68^{\mathrm{stat}} \pm 3.40^{\mathrm{syst}}) \times 10^{14}~\mathrm{M}_{\odot}$. Although the disturbed dynamical state of \moo\ is not striking from the visual inspection of the NIKA2 map at 150~GHz, this multi-wavelength analysis has shown that the precision of the mass measurement of this particular cluster is not limited by the noise properties in the \chandra\ and NIKA2 data but by our modeling of the ICM thermodynamic properties.\\

The knowledge of the systematic uncertainty on $M_{500}$ induced by an incorrect modeling of the ICM properties is essential to perform an accurate calibration of the mass-observable scaling relation. This analysis highlights the importance of the combination of SZ and X-ray observations to accurately measure the dynamical properties of high redshift galaxy clusters. Furthermore, it shows that a 1D modeling of the ICM is intrinsically unsuited to describe the true distributions of the thermodynamic properties of morphologically disturbed clusters. In this context, mapping the average value of the gas properties along the line of sight can provide new insights into the dynamical state of such unrelaxed systems.

\section{Spatially Resolved Maps of ICM Properties}\label{sec:moo_2d}

\subsection{Method}

\begin{figure*}[t]
\centering
\includegraphics[height=6.4cm]{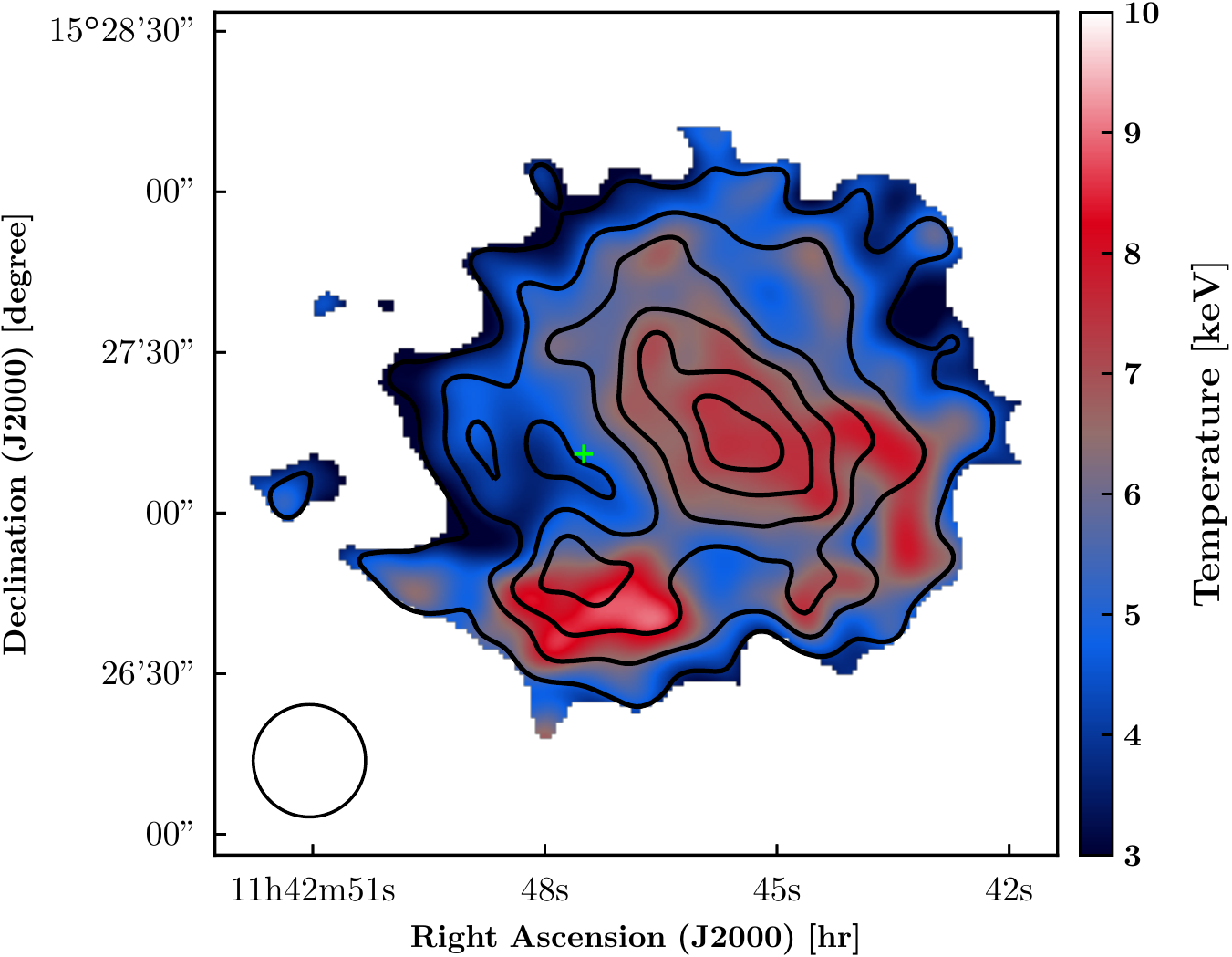}
\hspace{0.6cm}
\includegraphics[height=6.4cm]{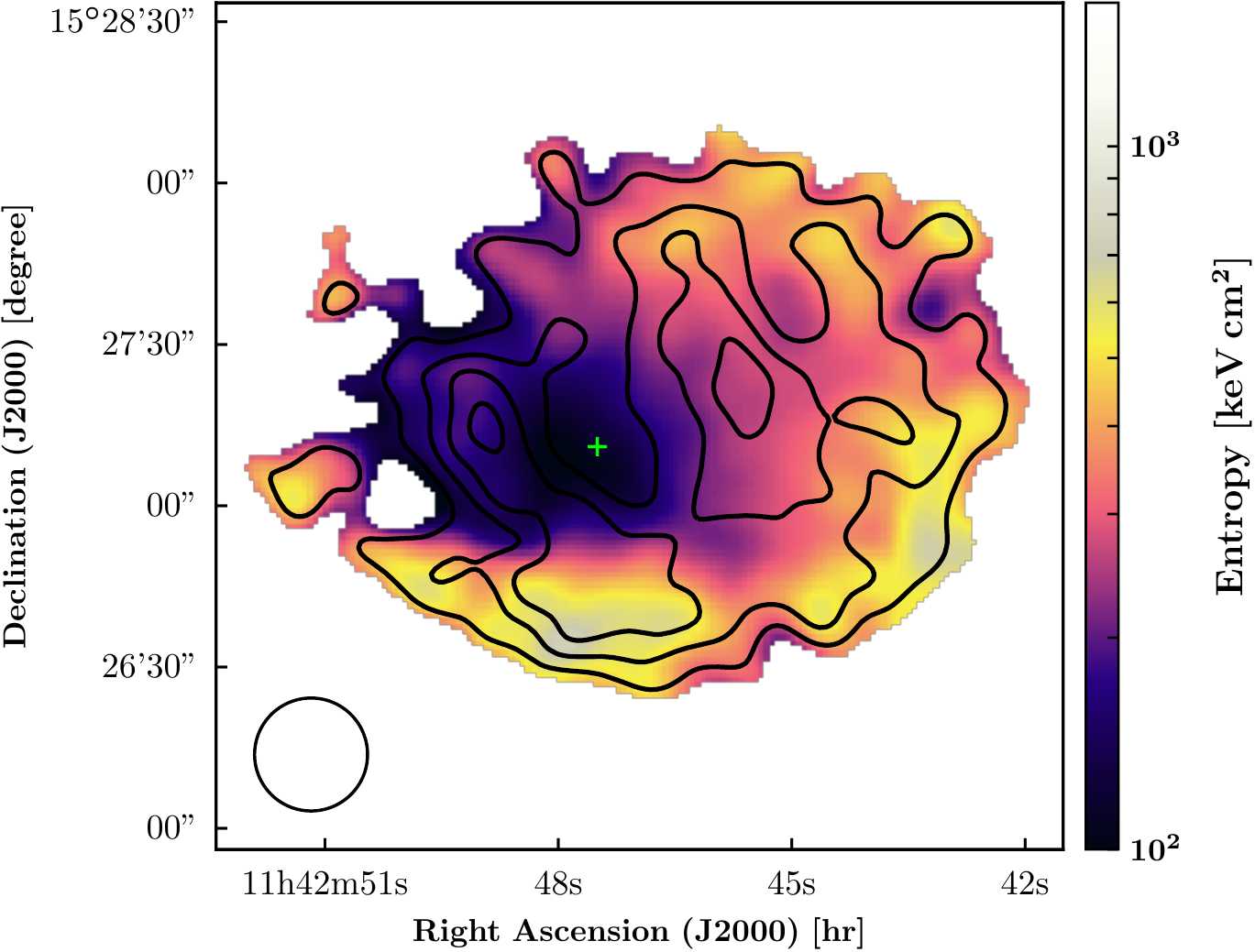}
\caption{{\footnotesize Temperature (left) and entropy (right) maps of \moo\ obtained from the combination of the pressure and density maps of the ICM measured with the NIKA2 and \chandra\ data respectively, see Eq. (\ref{eq:P_map}) and (\ref{eq:n_map}). The signal-to-noise (S/N) contours shown in black start at a value of $3\sigma$, increasing in steps of $1\sigma$. We only show the estimated temperature and entropy values in the regions where the S/N is higher than 2.}}
\label{fig:t_k_maps}
\end{figure*}

Mapping the ICM temperature in disturbed clusters using X-ray spectroscopic data has led to a better understanding of the complex  processes occurring in sloshing \citep[\emph{e.g.}][]{cal19} or merging \citep[\emph{e.g.}][]{mil09} systems. However, the realization of such maps usually requires thousands of X-ray counts per resolution element in order to have a precise measurement of the relative difference between the temperature estimates in each region of the map. Mapping the ICM temperature of \moo\ from its core up to a large fraction of $R_{500}$ based on X-ray spectroscopy measurements only would therefore require more that 1~Ms of exposure with \chandra\ or \xmm\ even with a very coarse binning of the map.\\

The different dependencies of the X-ray and SZ surface brightness with respect to the distribution of density and temperature along the line of sight allow us to combine the \chandra\ and NIKA2 data at the pixel level in order to map all the ICM thermodynamic properties of \moo\ without relying on X-ray spectroscopy. We follow the methodology described in detail in \cite{ada17b} in order to estimate the maps of the ICM pressure ($\bar{P}_e$) and density ($\bar{n}_e$) distributions averaged along the line of sight based on the NIKA2 and \chandra\ data:
\begin{equation}
\bar{P}_e(\alpha, \delta) = \frac{1}{\ell_{\mathrm{eff}}} \int P_e \, dl = \frac{m_e c^2}{\sigma_T} \frac{y_{tSZ}}{\ell_{\mathrm{eff}}}
\label{eq:P_map}
\end{equation}
\begin{equation}
\bar{n}_e(\alpha, \delta) = \frac{1}{\ell_{\mathrm{eff}}} \int n_e \, dl = \frac{1}{\sqrt{\ell_{\mathrm{eff}}}} \sqrt{\frac{4\pi \, (1+z)^4 \, S_X}{\Lambda(T_e,Z)}}
\label{eq:n_map}
\end{equation}
where $y_{tSZ}$ and $S_X$ are the NIKA2 Compton parameter map and \chandra\ X-ray surface brightness map, respectively, and $\ell_{\mathrm{eff}}$ is the map of the effective electron depth given by:
\begin{equation}
\ell_{\mathrm{eff}} = \frac{\left(\int n_e \, dl\right)^2}{\int n_e^2 \, dl}.
\end{equation}
This map provides an estimate of the line-of-sight extension of the ICM in each pixel. We compute two estimates of this map based on the ICM density profiles obtained using the X-ray peak and X-ray centroid as deprojection centers. Indeed, the uncertainty associated with the estimate of $\ell_{\mathrm{eff}}$ is dominated by the departure of the ICM geometry from the spherical model used in this analysis. We favor the X-ray peak instead of the X-ray centroid to compute the final maps $\bar{P}_e(\alpha, \delta)$ and $\bar{n}_e(\alpha, \delta)$ because our goal is to accurately map the ICM thermodynamic properties of \moo\ in the core area. However, we compute the uncertainties associated with these maps using both estimates of $\ell_{\mathrm{eff}}$. We measure a minimum effective electron depth of ${\sim}500$~kpc at the X-ray peak in the $\ell_{\mathrm{eff}}$ map obtained using the density profile centered on the X-ray peak. The cluster extent increases towards the outskirts and reaches a value of ${\sim}1400$~kpc at 1~arcmin from the X-ray peak.\\

The Compton parameter map measured by NIKA2 is deconvolved from the transfer function in order to minimize the bias induced by the large-scale filtering of the SZ signal. We estimate the standard deviation in each pixel of the pressure map by computing two samples of one thousand realizations of $\bar{P}_e$ using NIKA2 noise maps also deconvolved from the transfer function instead of $y_{tSZ}$ in Eq. (\ref{eq:P_map}) and the two realizations of $\ell_{\mathrm{eff}}$. We also take into account the effect of the point source subtraction in the Compton parameter map by simulating different realizations of $y_{tSZ}$ marginalizing over the radio source flux given the constraints obtained in Sect. \ref{sec:radio_pts}.\\
\begin{figure*}[t]
\centering
\includegraphics[height=6.4cm]{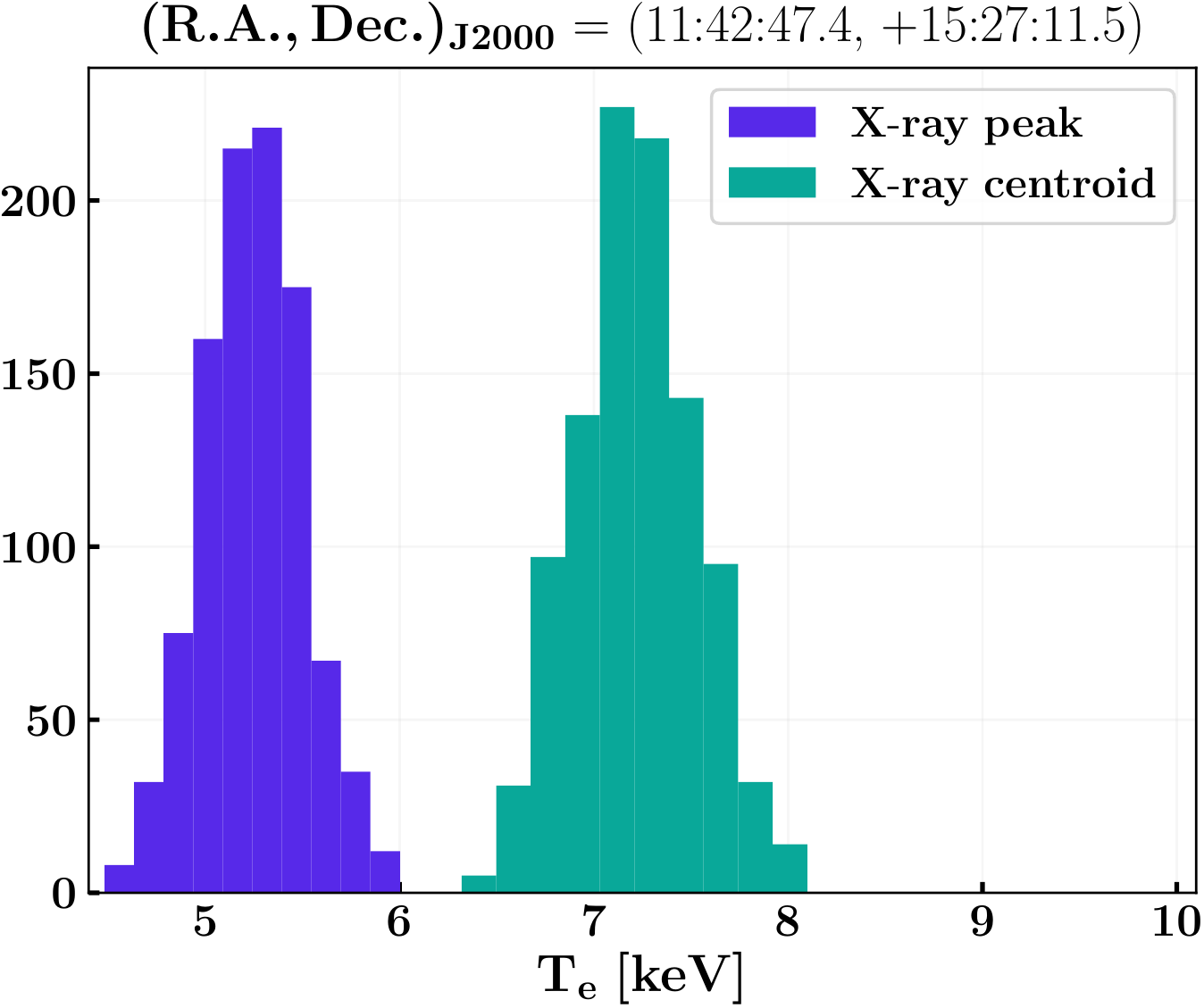}
\hspace{0.6cm}
\includegraphics[height=6.4cm]{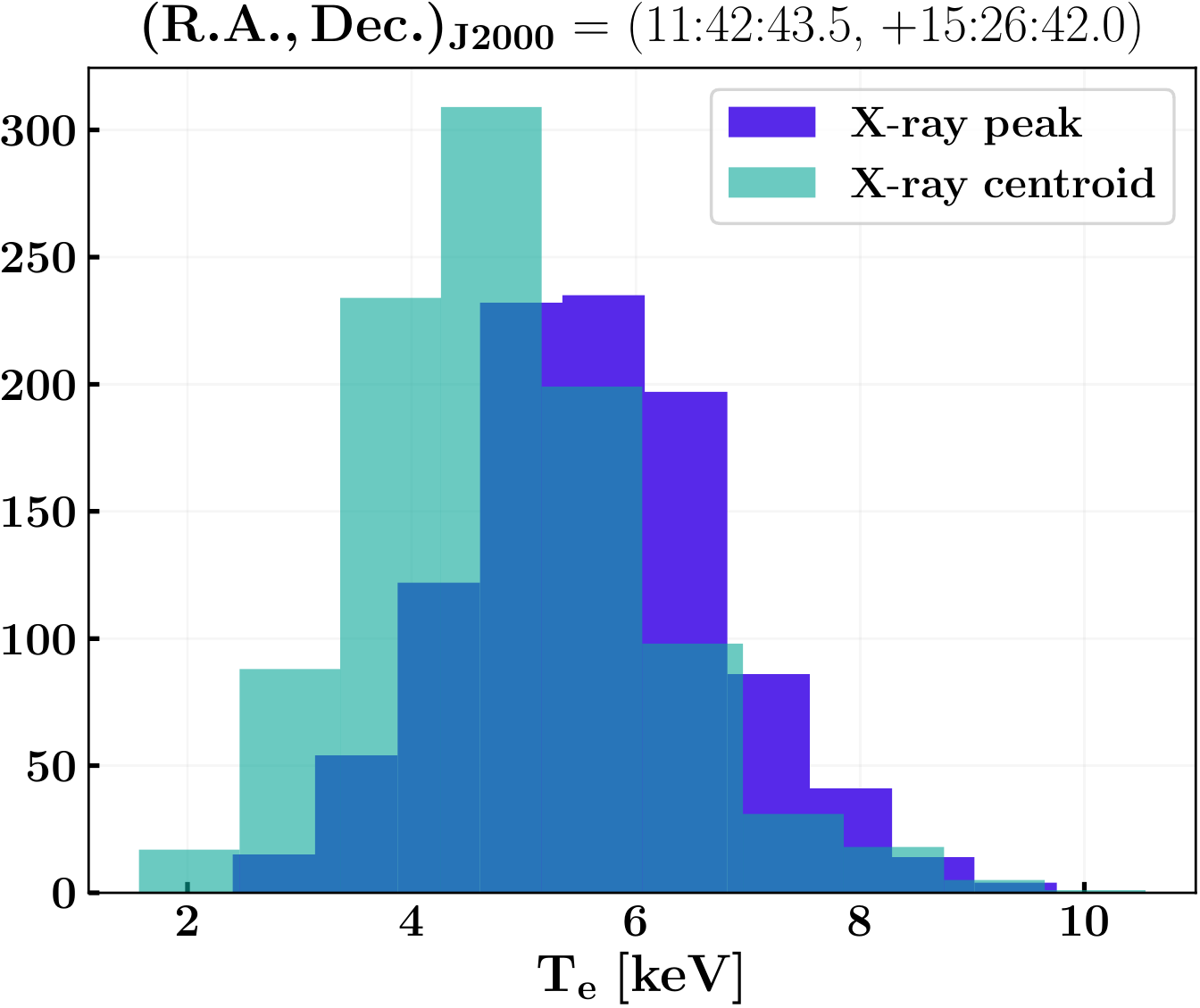}
\caption{{\footnotesize Distributions of the Monte Carlo realizations of the estimates of ICM temperature averaged along the line of sight measured in the core (left) and in the southwest (right) regions of \moo. The distributions in purple (resp. green) show the temperature values measured on the maps obtained by using the $\ell_{\mathrm{eff}}$ map estimated from the density profile centered on the X-ray peak (resp. X-ray centroid).}}
\label{fig:t_map_differences}
\end{figure*}

The \chandra\ map is corrected for vignetting and then processed through an adaptative filter using the CIAO \texttt{csmooth} routine. This allows us to produce a map where the minimum angular scales match the effective angular resolution of 20~arcsec adopted for the $\bar{P}_e$ map and where the minimum signal-to-noise ratio is equal to 3. Several realizations of this filtered map are made by taking into account the Poisson fluctuations of the X-ray signal and shuffling the pixel values of the X-ray background map before subtracting it from the \chandra\ surface brightness map. We use these $S_X$ realizations along with the two $\ell_{\mathrm{eff}}$ ones in order to compute two samples of one thousand realizations of $\bar{n}_e$ maps from Eq. \ref{eq:n_map}.\\

The temperature and entropy maps of \moo\ are estimated from the combination of the pressure and density map realizations:
\begin{equation}
k_B\bar{T}_e = \bar{P}_e / \bar{n}_e~~~~\mathrm{and}~~~~\bar{K}_e = \bar{P}_e / \bar{n}_e^{5/3}
\end{equation}
The results obtained by using the $\ell_{\mathrm{eff}}$ map computed with the ICM density profile centered on the X-ray peak are shown in both panels of Fig.~\ref{fig:t_k_maps}. We indicate the location of the BCG with a green cross. The signal-to-noise contours estimated from the knowledge of the temperature and entropy fluctuations in each pixel from the two thousand realizations of $\bar{P}_e$ and $\bar{n}_e$ maps are shown with black lines, with the lower contour at 3$\sigma$. We measure the spatial distribution of the ICM temperature and entropy up to a distance of $0.5R_{500}$ from the cluster centroid. We emphasize that the S/N on these temperature and entropy estimates take into account the effect of the differences between the two density profiles estimated in Sect. \ref{subsec:ne_prof} on the $\ell_{\mathrm{eff}}$ map. We show the distributions of the temperature values obtained in the core and in the southwest regions of the cluster from the Monte Carlo realizations of temperature maps computed by using the two estimates of $\ell_{\mathrm{eff}}$ in Fig.~\ref{fig:t_map_differences}. As the density estimates given at the location of the X-ray peak by the two profiles shown in Fig.~\ref{fig:1d_P_ne} are significantly different, the two distributions of the core temperature shown in the left panel are consequently different. This explains why the S/N on the core temperature and entropy values shown in Fig.~\ref{fig:t_k_maps} is much weaker than the one reached in the intermediate regions of the cluster where the temperature and entropy values obtained with the two estimates of $\ell_{\mathrm{eff}}$ are fully compatible (see right panel of Fig.~\ref{fig:t_map_differences}).

\subsection{Results}

The first noticeable feature in the temperature and entropy maps shown in Fig.~\ref{fig:t_k_maps} is the significant detection of a cool-core at the location of the BCG. The temperature measured at this location is at least two times lower than the one observed in the surrounding regions and the entropy is three times lower than the mean entropy value measured in the whole map. Furthermore, the core appears to be well-delimited. The core entropy averaged along the line of sight is enclosed between 100 and $200~\mathrm{keV\cdot cm^2}$ within a circular region extending over a ${\sim}120$~kpc radius from the BCG. We measure a significant increase of the ICM temperature to the west of the BCG. This feature is consistent with the merging scenario described in Sect. \ref{sec:moo_morphology} as we expect the gas within this region to be shock-heated by the infalling subcluster (B) from the northwest to the south-east of the main halo (A). The entropy measured at the location of the infalling galaxy group is slightly lower than the one observed at similar radii from the cluster BCG but in different directions. This could be a hint of the presence of a second core associated with this merging substructure. However this difference is not significant given the entropy fluctuations at this location.\\

An interesting feature identified in both maps is the presence of a high temperature and entropy region southward from the BCG location. While the relative difference between the temperature measured in this region and the one observed at the cluster centroid varies between all the realizations of $\bar{T}_e$, its absolute value is always twice higher than the one measured in the cool-core. The interpretation of this feature is quite challenging given its location with respect to the different structures in the galaxy distribution (see Sect. \ref{sec:moo_morphology}). There is however a hint of an additional peak in the galaxy distribution observed by \emph{Spitzer} to the east of the BCG (labeled C in Fig.~\ref{fig:rgb_map}). If this galaxy group is actually in a post-merger state and has undergone a head-on collision from the southwest regions to its current location, the ICM would have been shock-heated at the collision point during the merging process and the gas within the substructure would have been stripped away from its potential well. This would explain both the high temperature measured to the south of the X-ray peak and the low gas density observed to the east of the cluster at the location of the galaxy group (C). We note however that deeper observations would be needed to confirm this possible scenario. In particular, higher exposure X-ray observations would enable measuring the X-ray spectroscopic temperature in the region to the south of the X-ray peak. With our current X-ray data set, the spectrum extracted in this region is well fitted by an absorbed \texttt{APEC} model with $k_BT = 10$~keV ($\chi^2/ndf = 0.55$) as well as with $k_BT = 5$~keV ($\chi^2/ndf = 0.52$). An \xmm\ proposal led by I. Bartalucci, also an author on this paper, has been accepted to map the X-ray emission of \moo\ for a total of 105~ks. The larger effective area of \xmm\ combined with this higher exposure should enable measuring a precise value of the spectroscopic temperature in this region of the cluster.\\

We emphasize the complementarity between the results of the 1D analysis described in Sect. \ref{sec:moo_1d} and the maps of the ICM thermodynamic properties described in this section to explore the merger dynamics of \moo. To the best of our knowledge, this is the very first time that such a detailed analysis of the ICM has been achieved at $z>1$. This study paves the way to an in-depth characterization of the ICM properties of very high redshift galaxy clusters from a combination of spatially resolved X-ray and SZ observations.

\section{Future analyses}\label{sec:perspective}

The analysis presented in this paper has demonstrated how shallow X-ray observations of $z>1$ galaxy clusters can provide state-of-the-art descriptions of the ICM thermodynamic properties once combined with high angular resolution SZ observations. This represents a significant step forward in the characterization of the ICM dynamics at $z>1$ compared to previous X-ray analyses of SZ-selected clusters that relied on assumptions on the shape of the ICM profiles or on stacking analyses in order to constrain the ICM thermodynamic properties. We are planning to combine the outcome of this work with allocated \xmm\ data to investigate the large scale properties of the ICM. In particular, we will compare the \xmm\ spectroscopic temperature profile of this cluster with the mass weighted profiles shown in Fig.~\ref{fig:1d_T_K}. Mapping the X-ray signal at larger radii will also allow us to study the gas clumpiness as well as non-thermal pressure support around $R_{500}$.\\
\indent Furthermore, we intend to conduct such a joint analysis of SZ and X-ray data on a sample of high redshift clusters drawn from the MaDCoWS survey. This will allow us to investigate the redshift evolution of the ICM properties such as the entropy excess above the self-similar expectation and its link with merger activity, map the ICM temperature and study its variations given the presence of radio-loud AGN or merging substructures, and provide strong evolutionary constraints on the astrophysical processes involved during the most active part of cluster formation.\\
\indent Five other clusters at $0.93 \leqslant z \leqslant 1.75$ have already been observed by \chandra\ with at least 500 counts in the region of interest (Cycle 14-17-18, PIs: A. Stanford - M. Brodwin). Furthermore, a NIKA2 proposal has been accepted to carry out SZ mapping of these clusters and to reach similar signal-to-noise levels at 150~GHz as the ones measured for \moo. Once these observations are completed, we will be able to provide a description of the mean ICM thermodynamic properties in this sample of six clusters in a range of masses from $2.6$ to $6.1\times 10^{14}~\mathrm{M}_{\odot}$. The number of objects in this sample will of course need to be increased in order to have more statistically significant results on the redshift evolution of the mean ICM properties with respect to $z < 1$ clusters. Such a pilot study will greatly benefit the forthcoming optical-infrared surveys with the \emph{Euclid} satellite and the Large Synoptic Survey Telescope (LSST) \citep{ref10,euc19,lss09} by providing evolutionary constraints on cluster dynamics in the $1 < z < 2$ redshift range based on a multi-wavelength analysis combining SZ, X-ray, and infrared data.

\section{Summary and Conclusions}\label{sec:conclusions}

We have presented a joint analysis of spatially resolved X-ray and SZ observations of the $z=1.2$ cluster \moo, combining data sets obtained by \chandra\ and NIKA2. With a cleaned exposure of 46.2~ks, the number of X-ray counts in the cluster area is too low to deproject the ICM temperature profile based on spectroscopic data only. However, a joint analysis of the NIKA2 and \chandra\ data enables us to characterize the ICM thermodynamic properties with unprecedented precision at this redshift. Our main results are summarized below:
\begin{itemize}
\item[$\bullet$] We find a ${\sim}100$~kpc offset between the location of the X-ray peak and the X-ray large-scale centroid in the \chandra\ map. Furthermore, the SZ peak location in the NIKA2 map at 150~GHz is found at the frontier between the two peaks in the galaxy distribution mapped by the IRAC instrument on board \emph{Spitzer}. These two galaxy overdensities are located around the radio-loud BCG at the position of the X-ray peak and in the northwest of the cluster, respectively. The projected large-scale ICM morphology is found to be elliptical, with a main extension of the diffuse emission oriented along the right ascension axis, and with a slight tilt oriented towards the northwest. All these morphological features are consistent with a merging scenario where a subcluster located in the northwest is interacting with the main halo centered on the BCG.\\
\item[$\bullet$] The ICM density and pressure profiles are estimated from a deprojection of the \chandra\ and NIKA2 data, respectively. We identify a significant impact of the choice of deprojection center on the shape of the density profile estimated from the \chandra\ data. We conclude that using the X-ray peak is best suited for a measurement of the ICM density at the location of the BCG, whereas the X-ray large-scale centroid is more appropriate to estimate the global gas properties. However, this choice does not have a significant impact on the shape of the pressure profile estimated from the NIKA2 data because of the flat distribution of the SZ signal in the cluster core. Using the X-ray peak as deprojection center, we measure a cuspiness of the gas density of $\alpha = 0.95$ which indicates the presence of a cool-core.\\
\item[$\bullet$] We estimate the temperature, entropy, cooling time, and hydrostatic mass profiles without relying on X-ray spectroscopy by combining the \chandra\ density and NIKA2 pressure profiles. The complementarity between the SZ and X-ray data sets allows us to tightly constrain the shape of all these ICM profiles from the cluster core up to $0.7R_{500}$. The relative uncertainties on both the temperature and entropy profiles achieved by this joint analysis represent a large leap forward on the characterization of the ICM properties at $z>1$. The estimated temperature, entropy, and cooling time profiles are typical of cooling flow clusters. We measure a temperature drop from $9$~keV at $170$~kpc from the BCG to 4~keV in the core, a central entropy of ${\sim}25~\mathrm{keV\cdot cm^2}$, and  a central cooling time of $t_{cool} = (0.51\pm 0.13)$~Gyr. Combined with our measurement of the cuspiness and the identification of the BCG with a strong radio AGN at the location of the X-ray peak, these results lead us to the conclusion that \moo\ hosts a well-regulated cool-core very similar to those observed at low redshift.\\
\item[$\bullet$] We evaluated the systematic uncertainty caused by the spherical modeling of the ICM of this disturbed cluster on its hydrostatic mass estimate. We confirm the high mass of this system but find that the systematic uncertainty is significantly limiting the precision of our measurement: $M_{500} = (6.06 \pm 0.68^{\mathrm{stat}} \pm 3.40^{\mathrm{syst}}) \times 10^{14}~\mathrm{M}_{\odot}$. While this uncertainty could be lowered with a better modeling of the ICM, \emph{e.g.} with a triaxial model of the gas density and pressure distributions, we emphasize the importance of taking it into account when using such hydrostatic mass estimate to calibrate the mass-observable scaling relation used for cluster cosmology.\\
\item[$\bullet$] Taking advantage of the different dependence of the SZ and X-ray signals with respect to the line-of-sight distributions of the ICM density and temperature, we have produced temperature and entropy maps by combining the NIKA2 and \chandra\ data sets at the pixel level. We clearly identify the cool core within the ICM from a relatively low entropy value at the position of the BCG. In addition, we confirm the disturbed dynamical state of \moo\ from an analysis of the spatial variations in the plane of the sky of these ICM quantities averaged along the line-of-sight. In particular, we identify a high temperature region at the interface between the two peaks in the galaxy distribution suggesting that the gas has been shock-heated in this area. A more surprising, yet significant, feature revealed by these maps is the presence of a high-temperature and high-entropy region to the south of the BCG. This increase in temperature may have been induced by the galaxy group identified to the east of the BCG that might have already undergone a first passage with the main halo from south to east. This would explain both the lack of gas at the position of this galaxy group and the temperature excess in the collision region.
\end{itemize}
This work demonstrates the level of detailed analysis made possible by joint SZ/X-ray analyses on the characterization of the ICM of $z>1$ clusters, in particular on the dynamics of mergers and on the core properties. This study paves the way to a systematic analysis of galaxy clusters, during the most active part of their formation history, from the complementarity of current X-ray observatories and the new generation of high angular resolution millimeter instruments. This technique allows us to push the investigation of the ICM evolution to the $z > 1$ regime before the next generation of X-ray observatories such as \emph{Athena} and \emph{Lynx} come into play.

\section*{Acknowledgements}

We would like to thank the IRAM staff for their support during the campaigns. The NIKA dilution cryostat has been designed and built at the Institut N\'eel. In particular, we acknowledge the crucial contribution of the Cryogenics Group, and in particular Gregory Garde, Henri Rodenas, Jean Paul Leggeri, Philippe Camus. This work has been partially funded by the Foundation Nanoscience Grenoble and the LabEx FOCUS ANR-11-LABX-0013. This work is supported by the French National Research Agency under the contracts "MKIDS", "NIKA" and ANR-15-CE31-0017 and in the framework of the "Investissements d’avenir" program (ANR-15-IDEX-02). This work has benefited from the support of the European Research Council Advanced Grant ORISTARS under the European Union's Seventh Framework Programme (Grant Agreement no. 291294). We acknowledge fundings from the ENIGMASS French LabEx (R. A. and F. R.), the CNES post-doctoral fellowship program (R. A.), the CNES doctoral fellowship program (A. R.) and the FOCUS French LabEx doctoral fellowship program (A. R.). R.A. acknowledges support from Spanish Ministerio de Econom\'ia and Competitividad (MINECO) through grant number AYA2015-66211-C2-2. This work has benefited from the support of the European Research Council Advanced Grants ORISTARS and M2C under the European Unions Seventh Framework Programme (Grant Agreement Nos. 291294 and 340519). Support for this work was provided by NASA through SAO Award Number SV2-82023 issued by the Chandra X-Ray Observatory Center, which is operated by the Smithsonian Astrophysical Observatory for and on behalf of NASA under contract NAS8-03060. This work is based in part on observations made with the Spitzer Space Telescope, which is operated by the Jet Propulsion Laboratory, California Institute of Technology under a contract with NASA.

\end{document}